\documentclass[usenatbib]{mnras}
\usepackage{graphicx}
\usepackage{natbib}
\usepackage{latexsym}
\usepackage{amsmath}
\usepackage{setspace}
\usepackage{booktabs}
\usepackage{array, longtable}
\usepackage[bf, singlelinecheck=off]{caption}
\usepackage{subfig}
\usepackage[T1]{fontenc}
\usepackage{aecompl}

\newcommand{\etal}{et~al.} 
\newcommand{\ionhy}{H{\sc ii} }

\newcommand{\kms}{$\mbox{km~s}^{-1}$}

\newcommand{\degrees}{$^\circ$}

\newcommand{\specsfig}[1]        
{
   \begin{center}
     \begin{minipage}[t]{0.45\textwidth}
         \psfig{file=#1.eps,height=0.9\textwidth,angle=270}
     \end{minipage}
     \end{center}
 }

\newcommand{\specdfig}[2]        
{
   \begin{center}
     \begin{minipage}[t]{0.45\textwidth}
         \psfig{file=#1.eps,height=0.9\textwidth,angle=270}
     \end{minipage}
     \hfill
     \begin{minipage}[t]{0.45\textwidth}
         \psfig{file=#2.eps,height=0.9\textwidth,angle=270}
     \end{minipage}
   \end{center}
}

\begin{document}

\title[Class I and II Methanol Masers]{The relationship between Class~I and Class~II methanol masers at high angular resolution}
\author[McCarthy \etal]{T.\ P. McCarthy$^{1,2}$,\thanks{Email: tiegem@utas.edu.au} S.\ P. Ellingsen$^{1},$ M.\ A. Voronkov$^{2}$, G. Cim\`o$^{3,4}$ \\
  \\
  $^1$ School of Natural Sciences, University of Tasmania, Private Bag 37, Hobart, Tasmania 7001, Australia\\
  $^2$ Australia Telescope National Facility, CSIRO, PO Box 76, Epping, NSW 1710, Australia \\
  $^3$ Joint Institute for VLBI ERIC (JIVE), Postbus 2, 7990AA, Dwingeloo, The Netherlands\\
  $^4$ Netherlands Institute for Radio Astronomy (ASTRON), Postbus 2, 7990AA, Dwingeloo, The Netherlands}

 \maketitle

\begin{abstract}
We have used the Australia Telescope Compact Array (ATCA) to make the first high resolution observations of a large sample of class~I methanol masers in the 95-GHz ($8_0$--$7_1$A$^+$) transition.  The target sources consist of a statistically complete sample of 6.7-GHz class~II methanol masers with an associated 95-GHz class~I methanol maser, enabling a detailed study of the relationship between the two methanol maser classes at arcsecond angular resolution. These sources have been previously observed at high resolution in the 36- and 44-GHz transitions, allowing comparison between all three class~I maser transitions. In total, 172 95-GHz maser components were detected across the 32 target sources. We find that at high resolution, when considering matched maser components, a 3:1 flux density ratio is observed between the 95- and 44-GHz components, consistent with a number of previous lower angular resolution studies. The 95-GHz maser components appear to be preferentially located closer to the driving sources and this may indicate that this transition is more strongly inverted nearby to background continuum sources. We do not observe an elevated association rate between 95-GHz maser emission and more evolved sources, as indicated by the presence of 12.2-GHz class~II masers. We find that in the majority of cases where both class~I and class~II methanol emission is observed, some component of the class~I emission is associated with a likely outflow candidate.

\end{abstract}


\section{Introduction}

Interstellar masers are one of the most readily observable signposts of young high-mass star formation regions as strong emission is commonly  observed towards them from a range of molecular species and transitions.  The most commonly observed maser species in star formation regions are from the OH, water and methanol molecules, with the latter having the richest centimetre wavelength spectrum of the group.  Methanol is an asymmetric top molecule with hindered internal rotation around the C--O bond and it is these characteristics which produce a large number of near-degenerate rotational energy levels.  Methanol maser transitions are empirically divided into two classes, known as class I and class II.  The class II masers are the better studied methanol masers, with more than 1200 sources having been observed in the 6.7-GHz transition throughout the Milky Way \citep[e.g.][]{Caswell+10,Caswell+11,Breen+15, Green+10, Green+12a, Green+17} and more than 20 different transitions having been detected \citep[see][and references therein]{Ellingsen+12}.  Class~II transitions are closely associated with main-line OH masers, water masers and are found in compact clusters \citep[typical linear scale of less than 0.03~pc][]{Caswell97} close to the infrared sources which mark the location of a young, high-mass star. Class~I methanol masers have historically been harder to study as the strongest common transitions are at frequencies of 36- and 44-GHz which are accessible to less telescopes and furthermore, until recently there was no reliable method for targeting searches for these masers.  The class~I methanol masers are often observed to be distributed over scales of 0.1 -- 1~pc and are further offset from high-mass stars than class II masers \citep[e.g.][]{Kurtz+04, Voronkov+06}.  The empirically observed difference between the two classes of methanol masers is known to be caused by different pumping mechanisms \citep{Cragg+92, Voronkov+05b}, with the class I methanol masers being collisionally pumped, while the class~II masers are radiatively pumped \citep{Sobolev+97a,Cragg+05}. Class I masers are typically associated with the interface regions of molecular outflows \citep[e.g][]{Kurtz+04, Cyganowski+12, Voronkov+06, Voronkov+14}. However, these masers may also be associated with shocks resulting from expansion of \ionhy regions, EGOs or cloud-cloud interactions \citep{Sjouwerman+10, Cyganowski+09, Cyganowski+12, Voronkov+10a, Voronkov+14}.

Early studies of class~I and class~II methanol masers led to suggestions that they may be associated with different types of sources, however, this turned out to be an observational bias in that some of the best studied class~I (e.g. Orion KL) and class~II (e.g. W3(OH)) maser sources show no emission, or only very weak emission in the other class.  \citet{Slysh+94} made a search for 44-GHz class I masers towards a sample of star formation regions known to host either OH maser and/or 6.7-GHz class II methanol masers.  They detected many new class~I methanol masers in this search and claimed an anticorrelation between the velocity ranges and the peak intensity of the class~I and class~II methanol masers in the same sources.  \citet{Ellingsen05} searched for 95-GHz class~I methanol masers towards a statistically complete sample of class~II methanol masers, obtaining detections towards approximately 40 percent of the sample, however, \citeauthor{Ellingsen05} found no evidence for the anticorrelations between the two classes claimed by \citeauthor{Slysh+94}.

\begin{table*}
	\caption{The coordinates of the pointing centre for each of the 6.7-GHz class II maser sites observed, along with on-source time and RMS noise for the observations. Distance estimates to the target sources along with their associated error (values in parenthesis are the negative error bars in cases where errors are asymmetrical) are also included and drawn from \citet{Green+McClure11}. Three sources (G\,331.13$-$0.24, G\,333.128$-$0.44 and G\,333.466$-$0.16) did not have previously estimated linear distances, for these sources a near kinematic distance has been assumed using the same parameters and methodology presented in \citet{Green+McClure11} and their distances denoted with an asterix. The final column reports the linear offset between each class II pointing target and the closest detected 95-GHz maser component with velocity coincident to the velocity range of the class~II emission (for the estimated distance). Velocity range values for the class~II sources drawn from \citet{Caswell+11} and \citet{Green+12a}.}
	\begin{tabular}{lllcccccc} \hline
		\multicolumn{1}{c}{\bf Source} & \multicolumn{1}{c}{\bf RA}  & \multicolumn{1}{c}{\bf Dec} & \multicolumn{1}{c}{\bf Time Onsource} & \multicolumn{2}{c}{\bf RMS noise} & \multicolumn{1}{c}{\bf Dist.} & \multicolumn{1}{c}{\bf Dist. Error} & \multicolumn{1}{c}{\bf Class~I offset} \\
		\multicolumn{1}{c}{\bf name}     & \multicolumn{1}{c}{\bf (J2000)} & \multicolumn{1}{c}{\bf (J2000)} & & \multicolumn{1}{c}{\bf Cont} & \multicolumn{1}{c}{\bf Line} & & & \\
		& \multicolumn{1}{c}{\bf $h$~~~$m$~~~$s$}& \multicolumn{1}{c}{\bf $^\circ$~~~$\prime$~~~$\prime\prime$} & \multicolumn{1}{c}{ (min)} & \multicolumn{2}{c}{ (mJy beam$^{-1}$)} & (kpc) & (kpc) & (pc) \\   \hline \hline   
		G\,326.48+0.70 & 15 43 16.60 & $-$54 07 12.70 & 29.9 & 4.5 & 150 & 11.8 & 0.4 & 0.076 \\
		G\,326.64+0.61$^a$ & 15 44 32.90 & $-$54 05 28.60 & 29.9 & 2.0  & 60  & 2.3 & 0.4 & 0.037 \\
		G\,326.86$-$0.67$^a$ & 15 51 14.20 & $-$54 58 04.90 & 24.6 & 1.5  & 45  & 3.2 & 0.3(0.4) & $-$ \\
		G\,327.39+0.19 & 15 50 18.50 & $-$53 57 06.40 & 24.6 & 2.1 & 42  & 4.5 & 0.4(0.3) & 0.072 \\
		G\,327.63$-$0.11 & 15 52 50.20 & $-$54 03 00.70 & 24.6 & 1.9  & 45  & 8.9 & 0.4 & $-$\\
		G\,328.24$-$0.54$^a$ & 15 57 58.40 & $-$53 59 23.10 & 29.9 & 1.7  & 50  & 12.0 & 0.4 & 0.053 \\
		G\,328.25$-$0.53 & 15 57 59.80 & $-$53 58 00.90 & 29.9 & 2.3  & 45  & 2.6 & 0.4 & 0.007 \\
		G\,328.81+0.63$^a$ & 15 55 48.60 & $-$52 43 06.20 & 30.9 & 4.7  & 55  & 2.6 & 0.4 & 0.028 \\
		G\,329.03$-$0.20 & 16 00 31.80 & $-$53 12 49.70 & 24.8 & 2.5  & 50  & 12.0 & 0.4 & 0.112 \\
		G\,329.03$-$0.19 & 16 00 30.30 & $-$53 12 27.30 & 24.9 & 1.5  & 65  & 12.0 & 0.4 & 0.077 \\
		G\,329.07$-$0.30 & 16 01 09.90 & $-$53 16 02.70 & 21.4 & 1.5   & 48  & 11.8 & 0.4 & 0.312 \\
		G\,329.18$-$0.31 & 16 01 47.00 & $-$53 11 44.20 & 25.2 & 1.8  & 48  & 3.3 & 0.3 & 0.037 \\
		G\,329.47+0.50 & 15 59 40.70 & $-$52 23 27.70 & 24.6 & 1.4  & 44  & 10.8 & 0.3 & 0.026\\
		G\,331.13$-$0.24 & 16 10 59.70 & $-$51 50 22.70 & 25.2 & 1.7 & 45  & 5.2$^{*}$ & 0.3 & 0.032\\
		G\,331.34$-$0.34 & 16 12 26.50 & $-$51 46 16.90 & 24.6 & 2.2  & 45  & 3.9 & 0.3 & 0.048 \\
		G\,331.44$-$0.18$^a$ & 16 12 12.50 & $-$51 35 10.30 & 24.6 & 2.0  & 42  & 10.1 & 0.3 & 0.033\\
		G\,332.30$-$0.09 & 16 15 45.40 & $-$50 55 53.90 & 24.6 & 1.6  & 45  & 3.0 & 0.3(0.4) & 0.011\\
		G\,332.60$-$0.16 & 16 17 29.30 & $-$50 46 12.50 & 24.6 & 1.8  & 50  & 2.8 & 0.3(0.4) & 0.013\\
		G\,332.94$-$0.68$^a$ & 16 21 19.00 & $-$50 54 10.40 & 24.6 & 2.6  & 48  & 3.2 & 0.3(0.4) & $-$\\
		G\,332.96$-$0.67$^a$ & 16 21 22.90 & $-$50 52 58.70 & 24.6 & 1.7  & 45  & 2.8 & 0.3(0.4) & 0.006 \\
		G\,333.03$-$0.06$^a$ & 16 18 56.70 & $-$50 23 54.20 & 24.6 & 1.7  & 48  & 2.5 & 0.4 & 0.009 \\
		G\,333.07$-$0.44 & 16 20 49.00 & $-$50 38 40.70 & 24.6 & 1.8   & 45  & 3.3 & 1.5 & $-$\\
		G\,333.12$-$0.43 & 16 20 59.70 & $-$50 35 52.30 & 25.2 & 2.2  & 60  & 3.0 & 1.5 & 0.025\\
		G\,333.13$-$0.44 & 16 21 03.30 & $-$50 35 49.80 & 24.6 & 3.3 & 55  & 3.3$^{*}$ & 0.3(0.4) & 0.143\\
		G\,333.13$-$0.56$^a$ & 16 21 35.40 & $-$50 40 57.00 & 24.6 & 1.7  & 50  & 3.3 & 0.3 & 0.066\\
		G\,333.16$-$0.10$^a$ & 16 19 42.70 & $-$50 19 53.20 & 24.6 & 1.7   & 50  & 4.9 & 0.3 & 0.208\\
		G\,333.18$-$0.09 & 16 19 45.60 & $-$50 18 35.00 & 24.6 & 2.0   & 45  & 4.6 & 0.3 & 0.032\\
		G\,333.23$-$0.06 & 16 19 51.30 & $-$50 15 14.10 & 24.6 & 1.7 & 400 & 4.6 & 0.3 & 0.035\\
		G\,333.32+0.10 & 16 19 29.01 & $-$50 04 41.50 & 24.6 & 1.7  & 50  & 12.3 & 0.4(0.3) & 0.099\\
		G\,333.47$-$0.16 & 16 21 20.20 & $-$50 09 48.60 & 24.9 & 3.0  & 60  & 3.2$^{*}$ & 0.3(0.4) & 0.030\\
		G\,333.56$-$0.02 & 16 21 08.80 & $-$49 59 28.30 & 24.9 & 2.3  & 80  & 12.6 & 0.4 & 0.093\\
		G\,335.06$-$0.42$^a$ & 16 29 23.10 & $-$49 12 27.30 & 24.9 & 1.8  & 55  & 2.4 & 0.4 & 0.028\\
		&                    &                    &    &    &   & &   &\\ \hline
	\end{tabular} \label{tab:observations}
	\begin{flushleft}
		Note: $^a$sources are not within the statistically complete sample of 6.7-GHz methanol masers \\
	\end{flushleft}	
\end{table*}

The observations of \citeauthor{Ellingsen05} were made with a single dish and so while they established the statistical relationship between the fraction of class~II methanol maser sources which also host a class~I methanol maser nearby, they did not show whether the two classes of methanol maser are driven by the same young high-mass star, nor the detailed spatial relationship between them.  Here we report interferometric observations made with the Australia Telescope Compact Array of the 95-GHz class~I methanol masers towards the sources detected by \citet{Ellingsen05}.  The locations of the class~II methanol masers in all of these sources are known to sub-arcsecond precision from previous observations \citep{Caswell+11,Green+12a} and so we have the opportunity to make the first study of the relationship between class~I and class~II methanol masers at high resolution in a statistically complete sample.

\section{Observations} \label{sec:observations}

The observations were undertaken with the Australia Telescope Compact Array (ATCA) between 2004 September 24 and 26 (project code C1273).  The array was in the H75 configuration, with minimum and maximum baselines of 30.6m and 89.2m respectively, and the synthesised beam width for the observations at 95-GHz was approximately 6 $\times$ 5 arcseconds.  The correlator was configured with 256 spectral channels across a 32-MHz bandwidth. At an observing frequency of 95-GHz this corresponds to a total velocity range of 100~\kms\/ with a velocity resolution of 0.47~\kms\/ for uniform weighting of the lag function (a channel width of 0.4~\kms).

The target sources were the detected 95-GHz class I methanol masers from \citet{Ellingsen05}. \citeauthor{Ellingsen05} used the Mopra telescope to search for 95-GHz class I methanol masers towards a statistically complete sample of 6.7-GHz methanol masers in the Galactic longitude range $l = 325\text{\degrees}-335\text{\degrees}$ ; $b = \pm0.53\text{\degrees}$. We observed every source within the aforementioned Galactic longitude range that were observed to have 95-GHz emission in \citet{Ellingsen05}. Not every source observed by \citeauthor{Ellingsen05} were members of their statistically complete sample of 6.7-GHz methanol masers. Therefore, 11 of the 32 sources reported here are not within the statistically complete sample due to them falling outside the coordinate/LSR velocity range or being excluded for reasons described in \citet{Ellingsen05}. These 11 sources have been clearly noted in the observation tables, and are not included in analysis where a statistically complete sample is described. 

The data were reduced with {\sc miriad} using the standard techniques for ATCA 3mm spectral line observations. Amplitude calibration was with respect to Uranus and PKS\,B1921-293 was observed as the bandpass calibrator.  The data were corrected for atmospheric opacity and the absolute flux density calibration is estimated to be accurate to 20 percent.  This was estimated by comparing the measured flux density of the bandpass and phase calibrator sources over each of the three days of the observations.  The observing strategy interleaved 5 minutes onsource for each of the maser targets with 2.5 minute observations of a phase calibrator, alternating between two different phase calibrator sources 1600-44 and 1613-586. The data were self-calibrated (phase-only) using the emission from the brightest 95-GHz methanol maser in each pointing.  After self-calibration we used continuum subtraction (modelled using the spectral channels without maser emission) to isolate the spectral line and continuum emission components.

The 95-GHz maser emission was then subsequently imaged with a velocity plane width of 0.25~\kms. Imaging at higher than the spectral resolution of our observations does not affect our results due to the restrictions we place on determination of individual components. The pointing centre for each of the observations is listed in Table~\ref{tab:observations}, along with information on the total time onsource for each source and the RMS noise for both the continuum data and a 0.25~\kms\/ spectral resolution image cube.  We adopted a rest frequency of 95.169489-GHz for the $8_0$--$7_1$A$^+$ transition of methanol \citep{Muller+04}.

\begin{table*}
	\caption{Details of the continuum sources detected in 14 of the pointing targets. Including location of the peak emission, peak flux density, integrated flux density and whether any association is observed with either class II 6.7-GHz masers or detected 95-GHz masers.}
	\begin{tabular}{lccccccc} \hline
		\multicolumn{1}{c}{\bf Source} & \multicolumn{1}{c}{\bf RA}  & \multicolumn{1}{c}{\bf Dec} & \multicolumn{1}{c}{\bf Peak} & \multicolumn{1}{c}{\bf RMS noise} & \multicolumn{1}{c}{\bf Integrated} & \multicolumn{2}{c}{\bf Associations} \\
		\multicolumn{1}{c}{\bf name} & \multicolumn{1}{c}{\bf (J2000)} & \multicolumn{1}{c}{\bf (J2000)} &  \multicolumn{1}{c}{\bf Flux Density} &  & \multicolumn{1}{c}{\bf Flux Density} & \multicolumn{1}{c}{\bf Class II} & \multicolumn{1}{c}{\bf Class I} \\
		& \multicolumn{1}{c}{\bf $h$~~~$m$~~~$s$}& \multicolumn{1}{c}{\bf $^\circ$~~~$\prime$~~~$\prime\prime$} & \multicolumn{1}{c}{\bf (mJy)} & \multicolumn{1}{c}{\bf (mJy)} & \multicolumn{1}{c}{\bf (mJy)} & &\\   \hline \hline  
		G\,326.48+0.70 & 15:43:16.48 & $-$54:07:14.0 & 94 & 4.3 & 105.2 & y & y \\
		G\,326.64+0.61$^a$ & 15:44:32.85 & $-$54:05:28.6 & 23.2 & 2.1 & 41.9  & y & y \\
		G\,326.86$-$0.67 & & & $<5$ & & & & \\
		G\,327.39+0.19 & & & $<7$ & & & & \\
		G\,327.62$-$0.11 & & & $<6$ & & & & \\
		G\,328.24$-$0.54$^a$ & 15:57:58.37 & $-$53:59:21.9 & 63.9 & 1.6 & 66.2  & y & y \\
		G\,328.25$-$0.53 & & & $<8$ & & & & \\
		G\,328.81+0.63$^a$ & 15:55:48.48 & $-$52:43:07.2 & 973.3 & 4.8 & 1007  & y & y \\
		G\,329.03$-$0.20 & & & $<9$ & & & & \\
		G\,329.03$-$0.19 & & & $<5$ & & & & \\
		G\,329.07$-$0.30 & 16:01:07.86 & $-$53:16:05.8 & 9.6 & 1.6 & 9.7 & n & n \\
		& 16:01:09.99 & $-$53:16:00.3 & 7.4 & 1.6 & 7.2  & y & n \\
		G\,329.18$-$0.31 & 16:01:47.99 & $-$53:11:58.6 & 9.4 & 1.9 & 11.6  & n & n \\
		& 16:01:47.22 & $-$53:11:41.3 & 9.8 & 1.9 & 10.0 & y & y \\
		G\,329.47+0.50 & & & $<4$ & & & & \\
		G\,331.13$-$0.24 & 16:10:59.67 & $-$51:50:23.8 & 122.7 & 1.7 & 147.2  & y & y \\
		G\,331.34$-$0.34 & & & $<7$ & & & & \\
		G\,331.44$-$0.18$^a$ & 16:12:12.75 & $-$51:35:10.4 & 26.1 & 2.0 & 28.1  & y & y \\
		G\,332.30$-$0.09 & 16:15:45.58 & $-$50:55:55.3 & 17.9 & 1.6 & 27.4  & y & y \\
		G\,332.60$-$0.16 & & & $<7$ & & & & \\
		G\,332.94$-$0.68$^a$ & & & $<9$ & & & & \\
		G\,332.96$-$0.67$^a$ & & & $<6$ & & & & \\
		G\,333.03$-$0.06$^a$ & & & $<6$ & & & & \\
		G\,333.07$-$0.44 & 16:20:48.88 & $-$50:38:39.6 & 11.6 & 1.8 & 30.6  & y & y \\
		G\,333.12$-$0.43 & 16:20:59.78 & $-$50:35:50.6 & 14.0 & 2.4 & 14.0 & y & y \\
		G\,333.13$-$0.44 & 16:21:02.46 & $-$50:35:57.1 & 42.3 & 3.6 & 48.1  & y & y \\
		G\,333.13$-$0.56$^a$ & & & $<6$ & & & & \\
		G\,333.16$-$0.10$^a$ & 16:19:42.43 & $-$50:19:53.3 & 16.0 & 1.8 & 30.7  & y & y \\
		G\,333.18$-$0.09 & & & $<7$ & & & & \\
		G\,333.23$-$0.06  & 16:19:51.08 & $-$50:15:13.9 & 10.7 & 1.9 & 11.6 & y & y \\
		G\,333.32+0.10 & & & $<5$ & & & & \\
		G\,333.47$-$0.16 & & & $<9$ & & & & \\
		G\,333.56$-$0.02 & & & $<8$ & & & & \\
		G\,335.06$-$0.42$^a$ & & & $<6$ & & & & \\
		&   &  &   &   &  & &  \\ \hline
	\end{tabular} \label{tab:continuum}
\begin{flushleft}
	Note: $^a$sources are not within the statistically complete sample of 6.7-GHz methanol masers
\end{flushleft}	
\end{table*}

\section{Results} \label{sec:results}

We detected 95-GHz methanol maser emission towards all 32 of our target sources. A continuum source was detected in 14 out of 32 target pointings. The majority of continuum sources detected were point sources, however, more extended continuum emission was observed in four cases (G\,326.641+0.61, G\,331.132-0.24, G\,333.068-0.44 and G\,333.163-0.10). Table~\ref{tab:continuum} contains details of the location and flux density of these continuum sources, along with information on whether the 6.7-GHz pointing target or 95-GHz class I masers were associated. An association was considered to exist between the 95-GHz continuum emission and the two maser species, if the maser components were within 3 seconds of arc from the peak continuum emission. 

In all observed sources the methanol maser emission was detected in many spectral channels. Emission in consecutive spectral channels is often at a similar location and is therefore considered a single component. In order to describe the maser emission as a series of ``spots" we have to define a set of criteria for when we consider emission adequately separated in space and velocity to be considered a unique maser component. Here we are considering emission regions to be independent if they are separated by at least 1.4 seconds of arc (upper $3\sigma$ positional error across all sources) or separated by at least 1 km\,s$^{-1}$ in LSR velocity. If the angular separation criteria was met, the component was considered unique ; If only the velocity separation criterion was met, the spectrum of the component was closely considered in order to determine that it was unique, and not the result of broad emission from another component it shares a position with. The 95-GHz components detected in the target sources are listed within Table~\ref{tab:component_table}. In total, 172 unique 95-GHz maser components were detected across our 32 class~II targets. The absolute position of these components has been listed, along with their corresponding alphabetised location, as defined by Voronkov et al. (2014).

\begin{figure*}
	\captionsetup[subfigure]{justification=centering}
	\begin{center}
		\begin{minipage}[t]{1.05\linewidth}
			\begin{center}
				\subfloat[G326.859-0.67]{\includegraphics[scale=0.45]{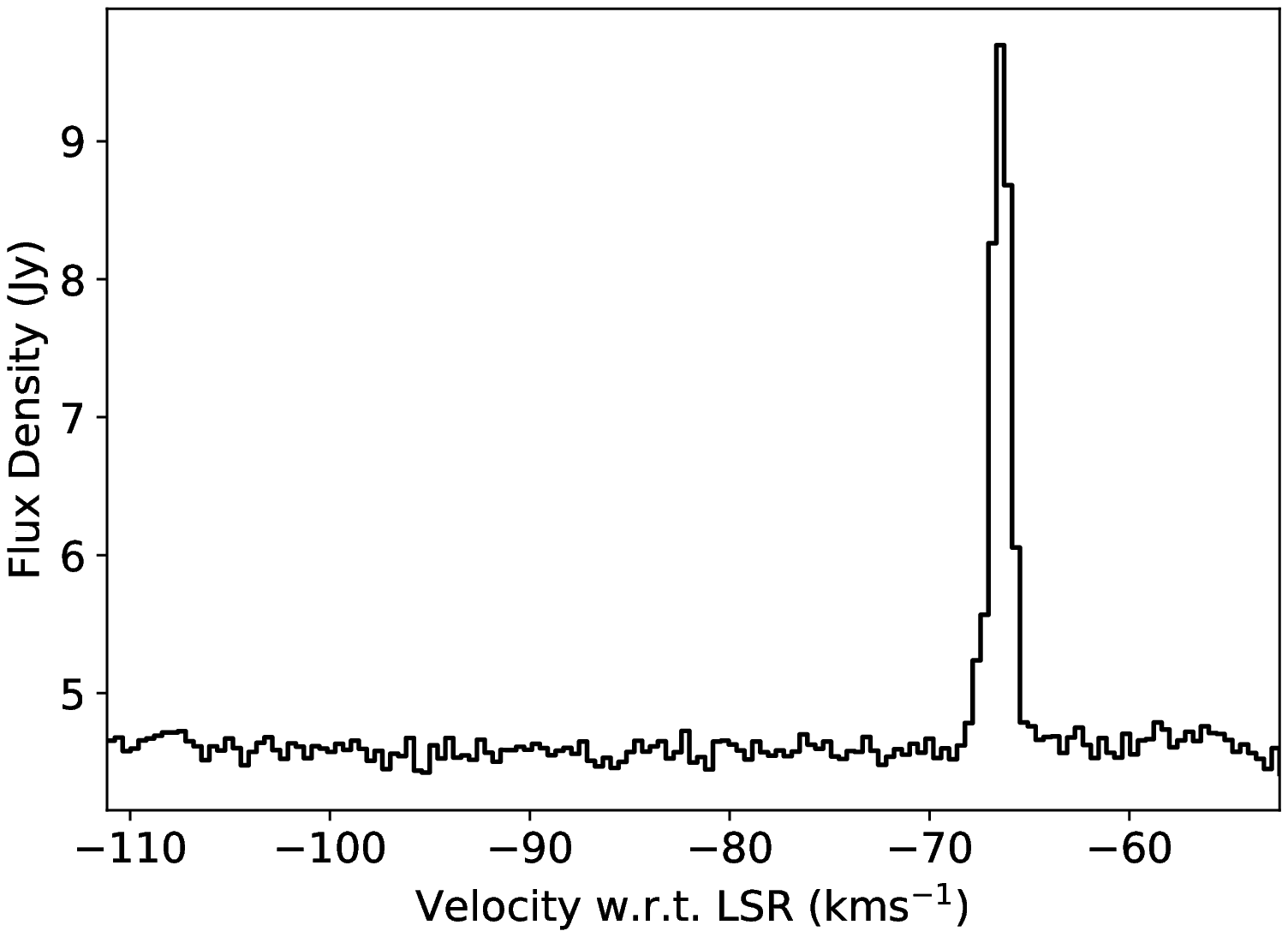}} 
				\hspace{4mm}
				\subfloat[G327.618-0.11]{\includegraphics[scale=0.45]{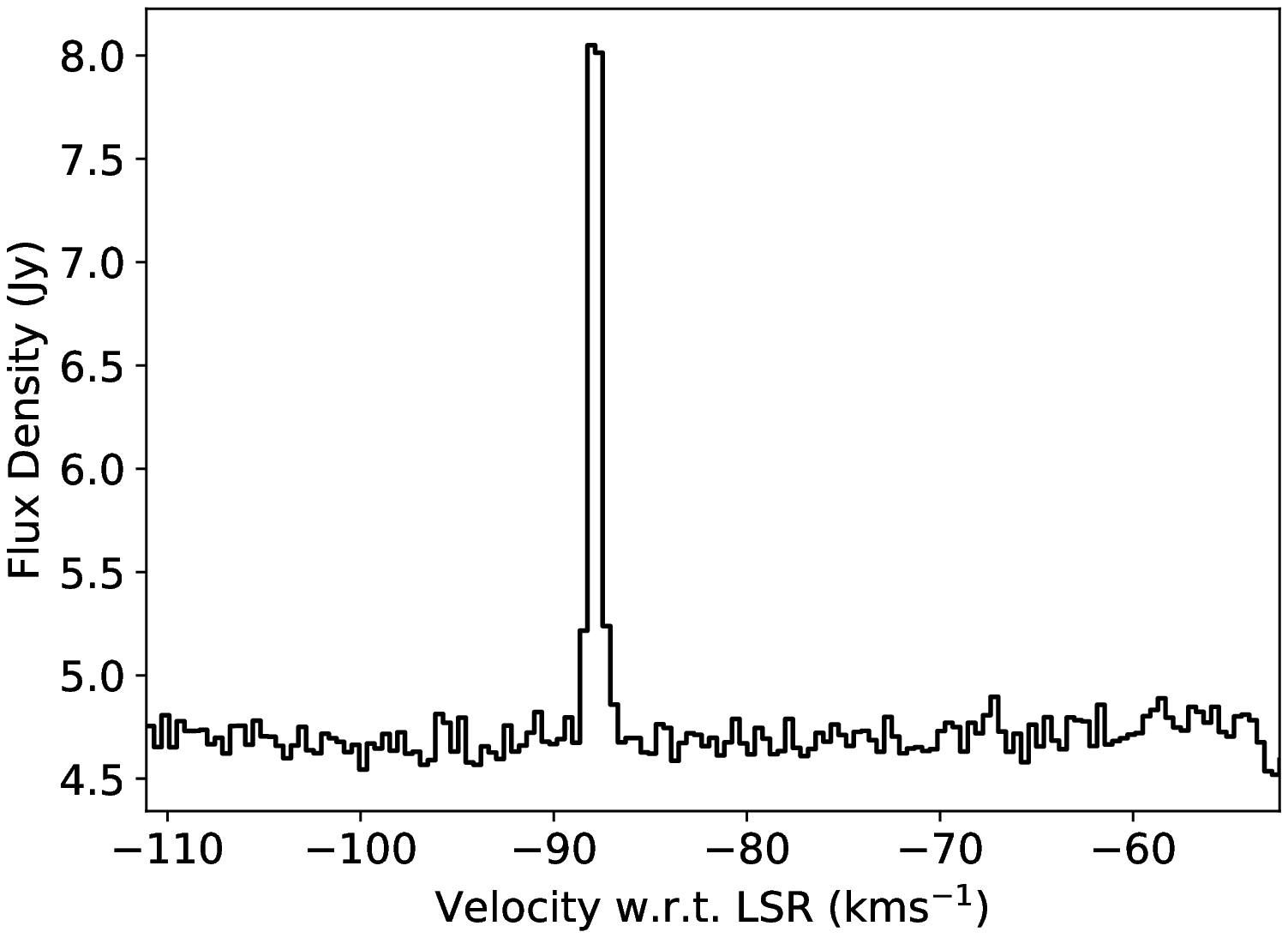}} \
				\subfloat[G329.066-0.30]{\includegraphics[scale=0.45]{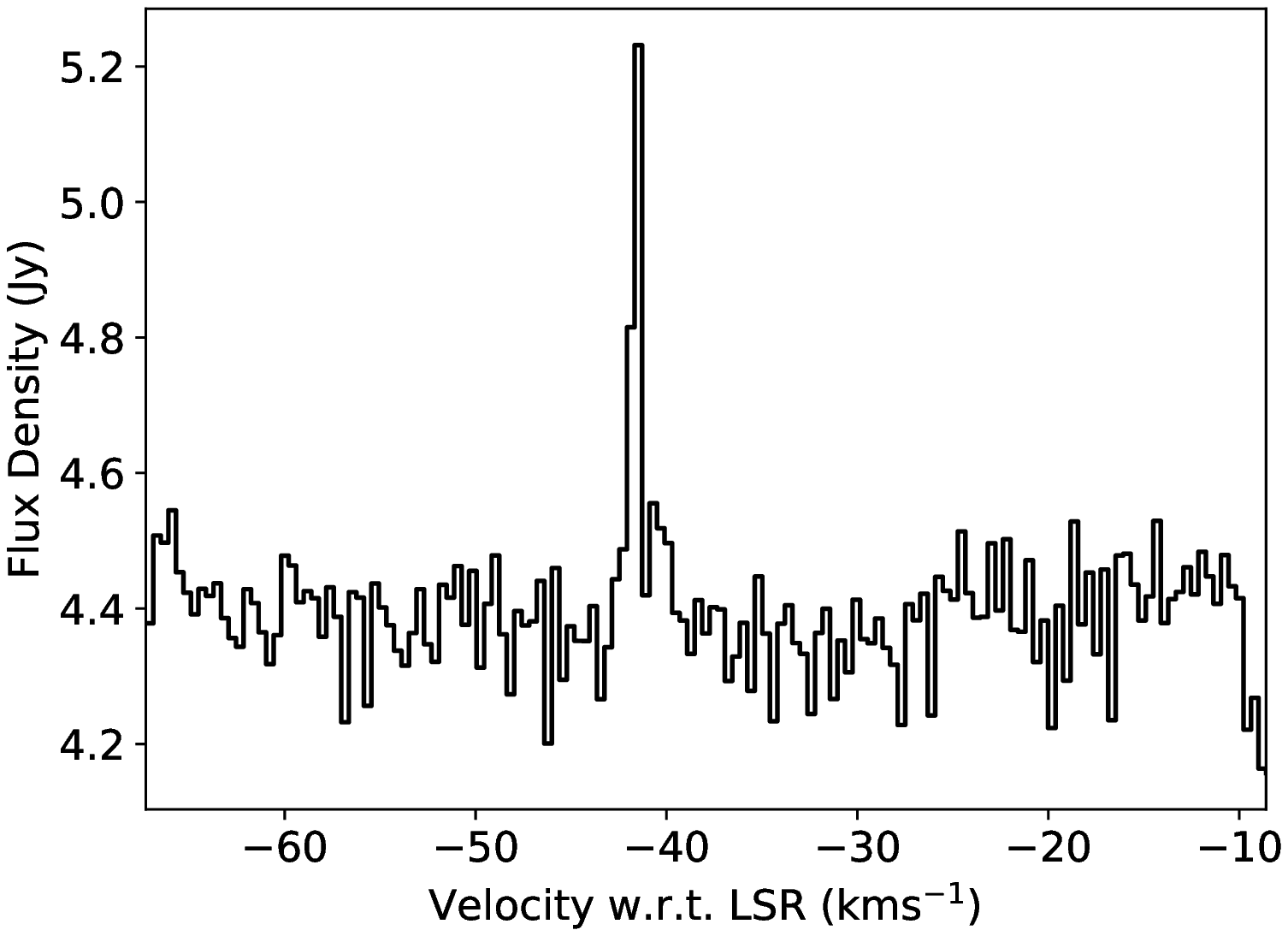}}
				\hspace{4mm} 
				\subfloat[G331.342-0.34]{\includegraphics[scale=0.45]{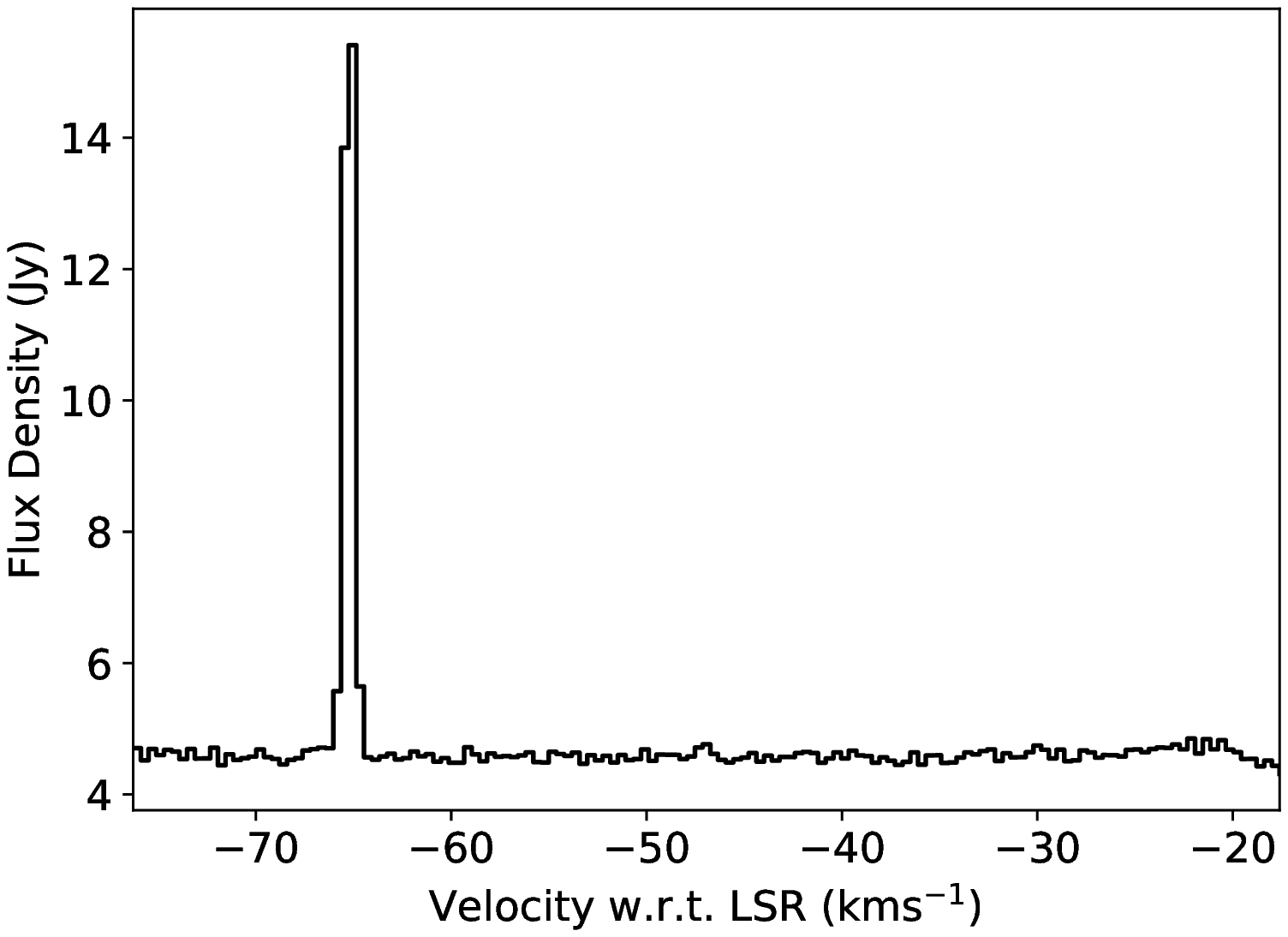}} \
				\subfloat[G332.604-0.16]{\includegraphics[scale=0.45]{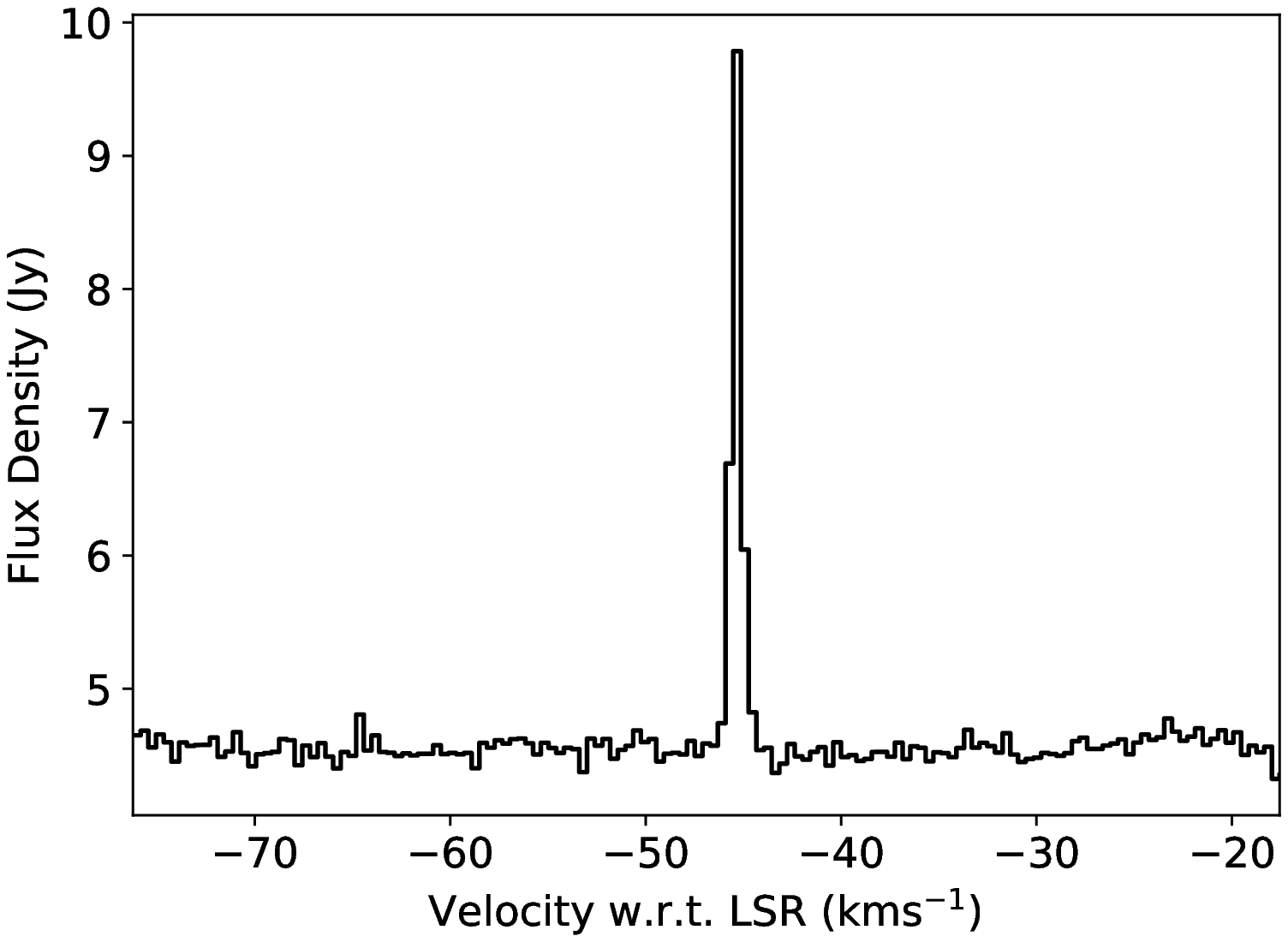}}
				\hspace{4mm} 
				\subfloat[G333.234-0.06]{\includegraphics[scale=0.45]{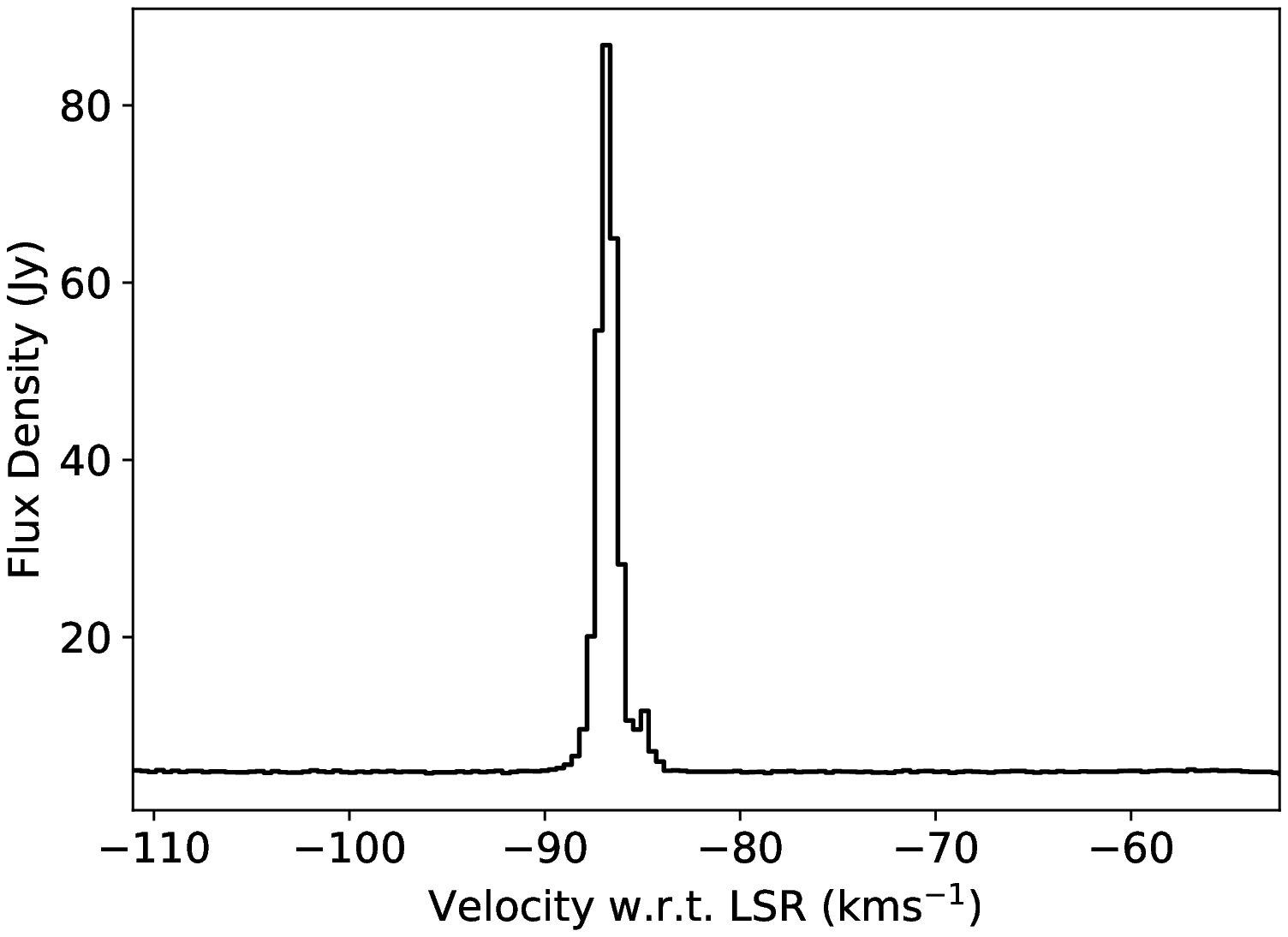}} \
			\end{center}
		\end{minipage}
	\end{center}
	\caption{Representative spectra of 95-GHz class I methanol masers from the current observations.  Spectra for the remaining sources in the sample are available online. The spectra were extracted from scalar averaging of the self-calibrated uv-data used to create spectral line image cubes.  Scalar averaged spectra allow us to see maser emission from all class~I locations in the primary beam, but have a positive offset baseline because of contributions from the system temperature of the receivers and any continuum emission in the primary beam.}
	\label{fig:example_spectra}
\end{figure*}

We aim to compare our observations to those presented in \citet{Voronkov+14}, who imaged 71 southern class I methanol maser sources at 36- and 44-GHz with the Australia Telescope Compact Array, with typical $1\sigma$ RMS noise levels of $\sim70$ and $\sim120$ mJy at 36- and 44-GHz respectively. \citeauthor{Voronkov+14} identified class I components that were roughly spatially associated and assigned them a letter name for ease of reference. We have used these same locations for grouping and referring to the components presented here. The 95-GHz components, along with the 36- and 44-GHz components from \citet{Voronkov+14} have been plotted on a series of field images (see Figure \ref{fig:fieldimages}). There are 30 field images total, covering the 32 targets sources from our observations, as two fields have multiple pointings (G329.03--0.20 and G333.13--0.44). In addition to these field images, Figure \ref{fig:example_spectra} shows scalar averaged spectra for all sources (except G328.25-0.53 where scalar averaging prevented emission from being clearly identifiable). These spectra all appear very similar to those presented in \citet{Ellingsen+05} for the same sources, albeit with generally higher SNR.

In all but three instances, locations containing 95-GHz components also harbour 44-GHz emission (as specified by \cite{Voronkov+14}). The exceptions occur in the sources G328.25$-$0.53 and G328.81+0.63, where the 95-GHz masers are located within $\sim4''$ of a class~II 6.7-GHz maser, but do not have an associated 44-GHz class I component.

A close association between 95- and 44-GHz methanol masers is expected because the two transitions belong to the same transition series ($\text{J}_0-(\text{J}-1)_1 A^+$) and there is a reported correlation between the observed flux density of the two transitions \citep{Valtts+00}. 

All, except one (G333.07$-$0.45), of our target sources had previously observed class~I maser emission (from the 36-GHz and 44-GHz transitions) reported by Voronkov et al. (2014). Due to the significantly smaller primary beam at 95-GHz ($\sim30"$), along with the expected flux density of the 95-GHz masers being half of that observed in associated 44-GHz masers, there are likely other 95-GHz components located outside the half-power range of our primary beam that we are unable to detect (specifically in regions with strong 44-GHz emission). \\



\subsection{Notes on individual sources} \label{sec:sourcenotes}

With the exception of G333.07$-$0.44, all the sources reported here have previous high resolution observations of the 36- and 44-GHz transitions by \citet{Voronkov+14}. In the comments below we have not repeated the information given by \citeauthor{Voronkov+14} relating to the general class I maser distribution, we only comment on the specifics of the new  95-GHz methanol maser information.

The field images presented in this study are created in the same way as those previously presented by Voronkov et al. (2014) and therefore any prior comment on the \textit{Spitzer} background images for individual sources will apply here. 

\subsubsection{G326.48+0.70}

\begin{figure*}
	\centering
	\includegraphics[height=0.40\textheight]{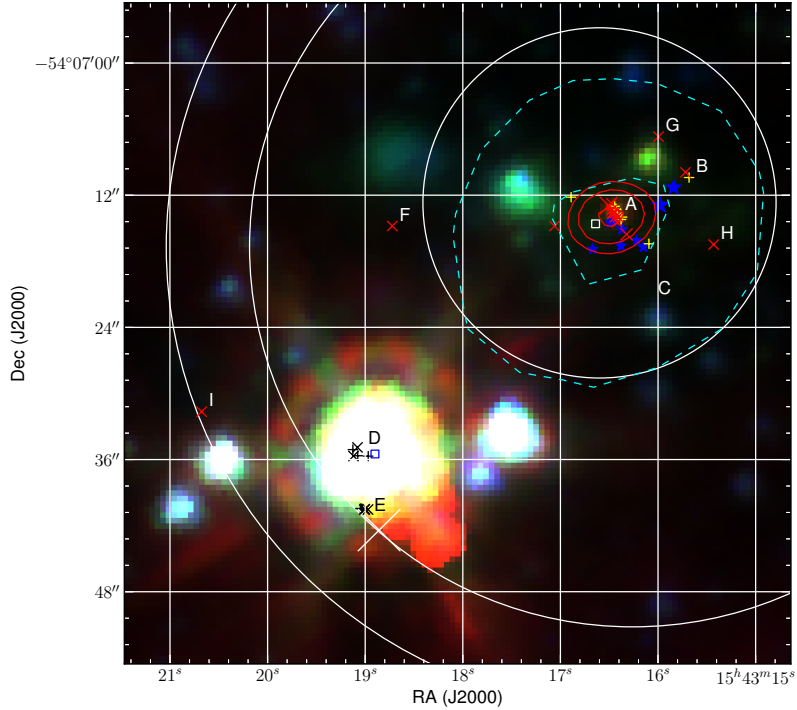}\\
	\caption{Field image for G326.48+0.70, field images for the remaining 29 class~I sources are available online. Four different symbols are projected on the \textit{Spitzer} background images, a square symbol represents a 6.7-GHz class II methanol maser, and the cross, plus and star symbols represent 36-,44- and 95-GHz class I maser emission respectively. Solid contours on the images represent 95-GHz continuum emission where applicable. Dashed contours represent the 50\% and 90\% levels of the ATLASGAL emission. The white circles, in order of increasing size, represent the full width at half maximum (FWHM) of the primary beam and are centred on the pointing direction for the 95-,44- and 36-GHz observations.}
	\label{fig:fieldimages}
\end{figure*}

The majority of 95-GHz methanol maser emission is associated with site A, close to the north-west class II 6.7-GHz maser. 95-GHz components trace the same linear structure, beginning at A and extending south-west, observed in the 36- and 44-GHz transitions. Within the FWHM of the 95-GHz primary beam there are two EGOs, both considered likely outflow candidates \citep{Cyganowski+08}. There is an EGO located at A (obscured by maser symbols at this location in Figure \ref{fig:fieldimages}), along with a larger region several arcsec to the east. Class I methanol maser emission is associated with both of these EGOs (only the western edge of the larger EGO), however, 95-GHz methanol maser emission is only spatially associated with the EGO at A. In addition to these two EGOs, there is another infrared source with 4.5-$\mu$m excess within the half-power range of our observations, located south-east of G. There is 3.5-mm continuum source detected at A, with peak intensity coincident with the densest cluster of class I methanol maser emission.  36- and 44-GHz emission has been detected in a second nearby 6.7-GHz maser, located in the south-east of this field and associated with a strong infrared source. However, this region is far beyond the half-power point of our 95-GHz pointing. A follow up observation of this south-eastern class II source would likely provide additional detections of 95-GHz class I methanol emission in this region.

\subsubsection{G326.64+0.61}

This 6.7-GHz target source is located on the edge of a large 8.0-$\mu$m filament (seen in the south-east corner of the field). Class I emission in this field is widely spread, however, the half-power range of our observations contains the majority of locations of previously detected methanol maser emission (A, B, D, E, G and J). 95-GHz maser emission is observed in two locations (B and D). A 3.5-mm continuum source is detected near D and E ($\sim3$ arcsec north-west of the 6.7-GHz maser), with the majority of class I maser emission at D associated with the southern edge. Additionally the 95-GHz masers at this location are arranged in a line extending from the edge of this continuum emission towards the south-west.

\subsubsection{G326.86$-$0.68}

Class I emission is relatively tightly clustered in this source, with locations B and A containing all the observed 95-GHz maser components. The class I emission at B is associated with a 6.7-GHz class II maser. East of the class II maser is a dark cloud, with class I emission at  B and C located at the interface of this region \citep{Peretto&09}. The 95-GHz masers in this source are located at the south-east edge of an EGO (likely outflow candidate according to \citet{Cyganowski+08}). 

\subsubsection{G327.39+0.20}

The methanol maser emission is separated into two separate locations (A and B), both with 6.7-GHz masers and associated with strong infrared sources. The 95-GHz pointing was centred on the westernmost 6.7-GHz maser (327.392+0.199) within this field. Although both locations are within the FWHM of the 95-GHz primary beam, 95-GHz maser emission is only observed at B. The infrared source associated with B also contains a 4.5-$\mu$m halo that has been classified as an EGO and considered a possible outflow candidate in \citet{Cyganowski+08}.

\subsubsection{G327.62$-$0.11}

The centre of this field contains a strong infrared source where the 6.7-GHz maser target and class I maser location A is situated. Only two 95-GHz components were observed in this source, one component tightly co-located with the 6.7-GHz maser, and one component at  B on the southern edge of the infrared source. Similar to what is observed in the other class I transitions, the 95-GHz maser emission at location A is considerably stronger than that observed in B \citep{Voronkov+14}.

\subsubsection{G328.24$-$0.55}

The 6.7-GHz pointing target is in the south-western corner of the field, with the majority of class I maser emission located nearby in A and B. The 95-GHz masers at A are loosely clustered, with the majority of these components being associated with other class I maser emission. The 6.7-GHz maser is a few arcsec south of A and associated with a 95-GHz component. A 3.5-mm continuum source is observed, co-spatially associated with the class II 6.7-GHz maser. The 95-GHz masers at A appear to follow some sort of arc that may be related to the 8.0-$\mu$m PDR situated north-east of these masers.

\subsubsection{G328.25$-$0.53}

The class II 6.7-GHz maser in this source is associated with a strong infrared source in the south-west of the field. The 36- and 44-GHz emission in this source is spread between 6 locations spanning over 1 arcmin squared, only one of which (A) is within the FWHM of the 95-GHz primary beam. The 95-GHz masers in this source are not associated with any of the locations where 36- and 44-GHz masers are detected, and instead they are associated with the 6.7-GHz maser target. This is one of only two cases across all of our sources where 95-GHz masers are observed offset from the other transitions of class~I emission. An EGO is also nearby to this class II source (likely outflow candidate according to \citet{Cyganowski+08}), with one 95-GHz component located at the interface of the EGO and the infrared source.

\subsubsection{G328.81+0.63}

Two strong class II 6.7-GHz masers, separated by $2.5''$, are projected onto a H \small{II} region in the centre of this field  \citep{Walsh+98,Green+12a}. The 95-GHz maser emission is spatially associated with the two 6.7-GHz masers at J. A 3.5-mm continuum source is detected at J, with peak emission at the location of the western 6.7-GHz maser. Interestingly there is no 44-GHz maser emission associated with the 95-GHz masers in this location, however, indicating the continuum source may be seeding the emission of these masers. This may occur due to the 95-GHz masers amplifying the underlying continuum emission, which may differ significantly between 7- and 3.5-mm, which boosts the expected ratio of 95- to 44-GHz flux densities at these projected locations.  95-GHz masers at B and C trace the southern edge of the diffuse 8.0-$\mu$m excess south of the H \small{II} region.

\subsubsection{G329.03$-$0.20}

This field is contains two separate 95-GHz pointings, one centred on the north-western 6.7-GHz maser 329.031-0.198 and the other on the south-eastern 6.7-GHz maser 329.029$-$0.205. Class I maser emission in the south-east pointing is organised in a linear structure (from F to H), extending north and south from the class II pointing target. The majority of the 95-GHz emission in the north-west pointing is clustered about the 6.7-GHz maser (location A). An EGO is present in this field (classified as likely outflow candidate by \citet{Cyganowski+08}), several arcsec south of A, with a single 36-GHz component located on its western edge.

\subsubsection{G329.07$-$0.31}

The class II 6.7-GHz maser target is located in the centre of a strong infrared source. An additional strong infrared source is on the eastern edge of the FWHM of the 95-GHz primary beam. Extended 4.5-$\mu$m emission is present near both of the infrared sources, with both being classified as possible outflow candidates by \citet{Cyganowski+08}. All detected 95-GHz maser components are located south-west of the 6.7-GHz maser, at the interface of the EGO between the western infrared source and location D. Interestingly we see the second case of a 95-GHz maser being offset from any previously observed class~I maser emission, with the maser spot associated with the western edge of the infrared source. There are two weak continuum sources detected in this field, the weaker continuum source is spatially associated with the 6.7-GHz maser, and the stronger (40 percent brighter) continuum source is located north-west of D, with no associated methanol emission or infrared source. 

\subsubsection{G329.18$-$0.31}

The 6.7-GHz maser target's location is projected onto a prominent EGO (likely outflow candidate according to \citep{Cyganowski+08}). The 95-GHz masers at E appear to trace the eastern boundary of the YSO and the EGO. The 95-GHz component at C is also located on the south-western edge of the EGO. Two weak continuum sources are detected in this region. the northern-most is positionally associated with the class I emission and 6.7-GHz maser at  E. The southern-most continuum source is outside the FWHM of the 95-GHz primary beam with no visible associations. Locations D and F both contain comparatively bright 44-GHz class I methanol masers and are within the half-power range of the 95-GHz primary beam, however, no 95-GHz emission was detected in either of these regions.

\subsubsection{G329.47+0.50}

The majority of class I maser emission is clustered around the 6.7-GHz maser in the centre of the field. This class II source is on the eastern edge of an EGO (classified as a likely outflow candidate by \citet{Cyganowski+08}). The majority of 95-GHz maser emission in this source is arranged in a line extending from the position of the 6.7-GHz maser towards the south-east.

\subsubsection{G331.13$-$0.24}

The 6.7-GHz source is located on the north-eastern edge of one of the northern-most H \small{II} regions in the field. This class II source is positioned at the interface between an EGO (likely outflow candidate according to \citet{Cyganowski+08}) and this H \small{II} region. 95-GHz masers are located at the three locations nearby to this H \small{II} region. The 95-GHz maser components are organised into a curve starting at the position of the 6.7-GHz maser, going through C and ending to the east at location B. A 3.5-mm continuum source is detected, encompassing the entire northern H \small{II} region, including the majority of 95-GHz emission.

\subsubsection{G331.34$-$0.35}

A strong infrared source dominates the centre of this compact class I maser. The 6.7-GHz maser is situated centrally to this infrared source. A class I masing component with emission from 36-, 44- and 95- GHz is situated at the interface between the strong infrared source and an EGO (classified as a possible outflow candidate by \citet{Cyganowski+08}). We observe another component at 95-GHz to the north-west, closer to the location of the 6.7-GHz maser.

\subsubsection{G331.44$-$0.19}

The 6.7-GHz class II target source is projected onto the edge of a compact H \small{II} region. 95-GHz masers are detected at all three locations (A, C and D) that fall within the half power point of the 95-GHz primary beam. A second 6.7-GHz maser is also located within the FWHM of the 95-GHz primary beam at C. 95-GHz emission in A and D trace a line across the H \small{II} region, connecting the two locations. The majority of 36- and 44-GHz components reported by \citet{Voronkov+14} form a line extending to the south-west from A through to E. Due to the line of 95-GHz components connecting D and A, class I maser emission from locations A through E may be all part of the same linear structure. A 3.5-mm continuum source is detected near D, with the emission between D and A tracing the outer edge of the emission.

\subsubsection{G332.30$-$0.09}

Two closely separated 6.7-GHz masers are within this field, the western 6.7-GHz maser (332.295$−$0.094, the pointing target for this source), is located at the southern edge of an EGO (possible outflow candidate according to \citet{Cyganowski+08}), and the second maser (332.296$−$0.094) is approximately $3''$ to the east. All but one of the 95-GHz maser components in this source are clustered between the northern edge of a H \small{II} region and the EGO to the north-west, with the other component situated $\sim15$ arcsec west. The 95-GHz components close to the eastern 6.7-GHz maser appear to be systematically offset by $\sim1$ arcsec south of the nearby 44-GHz maser components. An elongated 3.5-mm continuum source is detected at the location of the two 6.7-GHz masers, and encompasses all the 95-GHz maser emission at A.

\subsubsection{G332.60$-$0.17}

There are two sites of class I methanol emission within this source. The 6.7-GHz maser is associated with the southern class I emission and a faint EGO (classified as likely outflow by \citet{Cyganowski+08}). Both 95-GHz maser components are situated on the western edge of this EGO, nearby (within 2 arcsec) to the class II maser.

\subsubsection{G332.94$-$0.69}

The class II 6.7-GHz maser is located on the eastern edge of an infrared source, surrounded by an EGO (classified as likely outflow candidate by \citet{Cyganowski+08}) extending towards the south-east. All 95-GHz masers are situated at location A within the south-eastern region of the EGO. No 95-GHz maser emission is detected in location B, despite the relative brightness of the detected 44-GHz components.

\subsubsection{G332.96$-$0.68}

Location A, at the northern edge of a infrared point source with associated EGO (likely outflow candidate according to \citet{Cyganowski+08}), contains the majority of class I emission in this source. The EGO extends to the south-east, ending near B. The 6.7-GHz target maser is projected on this infrared point source, with the bulk of the observed 95-GHz emission clustered closely associated.

\subsubsection{G333.03$-$0.06}

The 6.7-GHz maser along with class I emission in this source is tightly clustered between a strong infrared source and a region of 4.5-$\mu$m excess (not classified as EGO by \citet{Cyganowski+08}). The 95-GHz masers here have a close association with the 44-GHz maser components.

\subsubsection{G333.07$-$0.45}

This is the source with no previous observation made by \citet{Voronkov+14}. The 6.7-GHz maser pointing target is projected onto the centre of a bright infrared region, surrounded by a region of 4.5-$\mu$m excess. There are two detected 95-GHz masers in this source, both offset to the north-west of the class II maser. An elongated 3.5-mm continuum source is detected, with peak emission co-located with the 6.7-GHz masers and an extension towards the south-east.

\subsubsection{G333.13$-$0.44}

Four 6.7-GHz class II masers are present in this field, with two separate 95-GHz pointings covering the densest regions of class I emission (targets 333.128-0.440 in the east and 333.121-0.434 in the west). The half-power range of the eastern pointing contains four distinct regions of class I emission (A, D, E and F), with a secondary 6.7-GHz maser associated with a strong infrared source at A. All 95-GHz masers from the eastern pointing form a line, starting at the position of the 6.7-GHz maser at A and extending towards the north-west. A 3.5-mm continuum source is also detected a few arcseconds south-west of this secondary 6.7-GHz maser.

The western pointing contains a single compact region of class I emission, clustered about the 6.7-GHz maser target (B), and located near the edge of a dark cloud. A 3.5-mm continuum source is also detected at this location, encompassing the 6.7-GHz maser and all but one of the 95-GHz maser components at B.

\subsubsection{G333.13$-$0.56}

The pointing target for this class I source was the southern-most of the two closely spaced 6.7-GHz masers, 333.128$-$0.560. 95-GHz masers are seen at three separate locations within this source, with the majority associated with an EGO (classified as possible outflow candidate by \citet{Cyganowski+08}) north-east of the northern 6.7-GHz maser (B). These 95-GHz components appear to be distributed in a line across the EGO. Bright 95-GHz emission is also observed at D, equidistant from both class II masers. Emission from all three different class I transitions are projected at this location (D), with the masers potentially being associated with a slight 4.5-$\mu$m excess. Voronkov et al. (2014) suggested that B and D may be connected with a curved 4.5-$\mu$m structure going through the nearby 6.7-GHz maser.

\subsubsection{G333.16$-$0.10}

6.7-GHz maser is associated with the eastern side of a strong infrared source, and H \small{II} region in the centre of the field. Class I maser emission is observed in two compact locations (A and B) in this source. The masers at B are close to the 6.7-GHz maser ($\sim2$ arcsec), whereas class I emission at A is offset to the south of the infrared source. An elongated 3.5-mm continuum source is detected, encompassing the infrared source and class I and class II emission at B. 

\subsubsection{G333.18$-$0.09}

The 6.7-GHz maser target is associated with an infrared point source surround by a prominent EGO (classified as likely outflow candidate by \citet{Cyganowski+08}) extending to the south-east. The 95-GHz maser components are spatially associated with the location of previously reported 36- and 44-GHz components at A along the northern edge of the EGO. 

\subsubsection{G333.23$-$0.06}

The south-eastern 6.7-GHz maser (333.234--0.062) is the pointing target of the 95-GHz observations in this source. A secondary 6.7-GHz maser (333.234--0.060) is located a few arcsec north-west, and is associated with a bright infrared source. 95-GHz emission is distributed between two locations, A and C. The former, is outside the FWHM of the 95-GHz primary beam, however, it contains a very bright 95-GHz component ($35.57$\,Jy, strong relative to the typical 95-GHz flux density values we observe). The 95-GHz masers at region C have a close spatial association with the position of the south-eastern 6.7-GHz maser. 3.5-mm continuum emission is detected, situated between the two 6.7-GHz masers on the south-eastern edge of the infrared source.

\subsubsection{G333.32+0.11}

The 6.7-GHz maser in this field is projected onto a strong infrared source, with an EGO (classified as likely outflow candidate by \citet{Cyganowski+08}) at the north-western edge. 95-GHz maser components at location A are projected across the infrared source, with one component of 95-GHz emission located significantly south-east of any previously reported class I emission. Interestingly in this source there does not appear to be a close spatial association between the 95-GHz components and the 36- and 44-GHz components. This may be due to evolution in the source, over the two and a half years between observations.

\subsubsection{G333.47$-$0.16}

The 6.7-GHz maser pointing target in this source is located within the southern half of an EGO (classified likely outflow candidate by \citep{Cyganowski+08}). Voronkov et al. (2014) notes a curved distribution of class I emission at locations A, B, D and F tracing a bow shock with the 6.7-GHz maser situated at the apex. The 95-GHz masers are consistent with the positions of the previously observed class I masers at A and B. However, at D one 95-GHz maser is located with the previously observed emission, and another is observed east of the 6.7-GHz maser at the edge of the EGO.  No 95-GHz masers are observed at F, possibly due to the 44-GHz maser here being relatively weak compared to other locations in this source. 95-GHz masers are located at E and C also, making it appear that these masers trace the entire eastern edge of the EGO.

\subsubsection{G333.56$-$0.02}

A mistake in the observations of this source caused the pointing to be $\sim18''$ north of the intended 6.7-GHz maser target. Therefore, all the class I emission is outside the FWHM of the 95-GHz primary beam to the south. All class I maser emission in this source is compactly clustered about this 6.7-GHz class II maser. One of the 95-GHz masers is offset to the north-east from the other class I emission.

\subsubsection{G335.06$-$0.43}

The class II maser is within the central region of an EGO (classified as a likely outflow candidate by \citet{Cyganowski+08}). The 95-GHz masers at B and D closely trace the southern edge of the EGO. \citeauthor{Voronkov+14} note that the class I emission at A appears to trace the northern edge of the EGO and the 95-GHz components at this location are consistent with this observation.

\setlength\LTleft{0pt}
\setlength\LTright{0pt}
\onecolumn\begin{longtable}{@{\extracolsep{\fill}}lccccc@{}}
	\caption{Detailed list of the 95\,GHz components detected in each maser site, including their alphabetical location as defined by Voronkov et al. (2014), positions and flux density. Source names have been converted from those of the individual pointings to those used by Voronkov et al. (2014); This results in two cases where two 95`GHz pointings have been combined into one source.} \label{tab:component_table} \\
	\hline     \multicolumn{1}{c}{\bf Source} & \multicolumn{1}{c}{\bf Location}  & \multicolumn{1}{c}{\bf LSR} & \multicolumn{1}{c}{\bf RA} & \multicolumn{1}{c}{\bf Dec} & \multicolumn{1}{c}{\bf Flux}  \\
	\multicolumn{1}{c}{\bf name}     & & \multicolumn{1}{c}{\bf Velocity} &  \multicolumn{1}{c}{\bf (J2000)} & \multicolumn{1}{c}{\bf (J2000)} & \multicolumn{1}{c}{\bf Density} \\
	& & \multicolumn{1}{c}{\bf (km\,s$^{-1}$)} & \multicolumn{1}{c}{\bf $h$~~~$m$~~~$s$} & \multicolumn{1}{c}{\bf $^\circ$~~~$\prime$~~~$\prime\prime$} & \multicolumn{1}{c}{\bf (Jy)}  \\
	\hline \hline \endfirsthead
	
	\multicolumn{6}{c}%
	{{\bfseries \tablename\ \thetable{} $-$- continued from previous page}} \\
	\hline      \multicolumn{1}{c}{\bf Source} & \multicolumn{1}{c}{\bf Location}  & \multicolumn{1}{c}{\bf LSR} & \multicolumn{1}{c}{\bf RA} & \multicolumn{1}{c}{\bf Dec} & \multicolumn{1}{c}{\bf Flux} \\
	\multicolumn{1}{c}{\bf name}     & & \multicolumn{1}{c}{\bf Velocity} &  \multicolumn{1}{c}{\bf (J2000)} & \multicolumn{1}{c}{\bf (J2000)} & \multicolumn{1}{c}{\bf Density} \\
	& & \multicolumn{1}{c}{ (km\,s$^{-1}$)} & \multicolumn{1}{c}{\bf $h$~~~$m$~~~$s$} & \multicolumn{1}{c}{\bf $^\circ$~~~$\prime$~~~$\prime\prime$} & \multicolumn{1}{c}{\bf (Jy)} \\ \hline \hline
	\endhead
	
	\hline 	\multicolumn{6}{c}{{Continued on next page}} \\
	\endfoot
	
	\hline \hline   \endlastfoot
	& & & & & \\
	G\,326.48+0.70 & A   & $-$47.0 & 15:43:16.67 & $-$54:07:16.9 & 0.31 \\
	& A   & $-$46.3 & 15:43:16.38 & $-$54:07:16.6 & 0.33 \\
	& A   & $-$45.2 & 15:43:16.49 & $-$54:07:14.5 & 0.85 \\
	& A   & $-$44.0 & 15:43:16.46 & $-$54:07:14.3 & 6.04 \\
	& A   & $-$41.7 & 15:43:16.37 & $-$54:07:15.1 & 8.18 \\
	& A   & $-$40.5 & 15:43:16.15 & $-$54:07:16.7 & 14.62 \\
	& A   & $-$37.7 & 15:43:16.22 & $-$54:07:16.1 & 1.96 \\
	& B   & $-$39.5 & 15:43:15.98 & $-$54:07:12.9 & 3.56 \\
	& B   & $-$39.3 & 15:43:15.84 & $-$54:07:11.3 & 4.31  \\
	G\,326.64+0.61$^a$ & B   & $-$43.0 & 15:44:33.60 & $-$54:05:20.9 & 0.50 \\
	& B   & $-$37.3 & 15:44:33.60 & $-$54:05:23.0 & 1.57 \\
	& D   & $-$41.0 & 15:44:32.96 & $-$54:05:31.0 & 0.99 \\
	& D   & $-$40.8 & 15:44:32.72 & $-$54:05:33.2 & 1.27  \\
	& D   & $-$40.0 & 15:44:32.95 & $-$54:05:30.8 & 3.17 \\
	& D   & $-$38.2 & 15:44:32.85 & $-$54:05:32.3 & 6.45 \\
	G\,326.86$-$0.67 & A   & $-$67.3 & 15:51:13.84 & $-$54:58:04.7 & 3.22 \\
	& A   & $-$66.7 & 15:51:13.73 & $-$54:58:05.0 & 8.31 \\
	& A   & $-$66.0 & 15:51:13.76 & $-$54:58:05.1 & 5.79 \\
	& A   & $-$65.5 & 15:51:13.91 & $-$54:58:05.2 & 1.44 \\
	& B   & $-$68.8 & 15:51:14.11 & $-$54:58:03.8 & 0.53 \\
	& B   & $-$67.8 & 15:51:14.05 & $-$54:58:03.9 & 2.04  \\
	G\,327.39+0.19 & B   & $-$88.8 & 15:50:18.53 & $-$53:57:06.3 & 3.30 \\
	& B   & $-$87.3 & 15:50:18.32 & $-$53:57:07.8 & 1.96  \\
	& B   & $-$86.5 & 15:50:18.13 & $-$53:57:07.4 & 0.57 \\
	G\,327.62$-$0.11 & A   & $-$88.0 & 15:52:50.20 & $-$54:03:00.9 & 6.78 \\
	& B   & $-$86.8 & 15:52:50.31 & $-$54:03:05.5 & 0.88  \\
	G\,328.24$-$0.54$^a$ & A   & $-$42.3 & 15:57:58.38 & $-$53:59:22.9 & 0.99 \\
	& A   & $-$41.7 & 15:57:58.34 & $-$53:59:20.3 & 1.12 \\
	& A   & $-$40.8 & 15:57:58.27 & $-$53:59:17.7 & 7.55 \\
	& A   & $-$40.0 & 15:57:58.39 & $-$53:59:20.5 & 0.42 \\
	& A   & $-$39.3 & 15:57:58.55 & $-$53:59:19.5 & 0.54 \\
	& A   & $-$37.7 & 15:57:58.66 & $-$53:59:19.8 & 0.42 \\
	& B   & $-$44.0 & 15:57:58.46 & $-$53:59:31.3 & 0.49 \\
	& B   & $-$43.2 & 15:57:58.28 & $-$53:59:31.3 & 1.86 \\
	G\,328.25$-$0.53 & -   & $-$44.0 & 15:57:59.93 & $-$53:57:57.4 & 0.53 \\
	& -   & $-$43.5 & 15:57:59.82 & $-$53:58:01.5 & 0.47  \\
	& -   & $-$41.7 & 15:57:59.81 & $-$53:58:00.5 & 1.67 \\
	G\,328.81+0.63 & A   & $-$42.0 & 15:55:50.23 & $-$52:43:21.7 & 4.72 \\
	& B   & $-$42.5 & 15:55:49.15 & $-$52:43:23.6 & 3.66 \\
	& C   & $-$41.0 & 15:55:48.25 & $-$52:43:19.8 & 2.46 \\
	& C   & $-$40.3 & 15:55:48.49 & $-$52:43:19.5 & 14.89 \\
	& J   & $-$44.7 & 15:55:48.48 & $-$52:43:06.5 & 1.22 \\
	& J   & $-$43.5 & 15:55:48.53 & $-$52:43:05.8 & 1.66 \\
	& J   & $-$38.0 & 15:55:48.84 & $-$52:43:07.1 & 0.50 \\
	G\,329.03$-$0.20$^{a\,b}$ & A   & $-$46.5 & 16:00:30.33 & $-$53:12:28.7 & 1.65 \\
	& F   & $-$43.2 & 16:00:31.98 & $-$53:12:45.7 & 18.87 \\
	& F   & $-$42.0 & 16:00:31.84 & $-$53:12:47.2 & 9.09 \\
	& G   & $-$41.2 & 16:00:31.90 & $-$53:12:55.8 & 3.67 \\
	& G   & $-$40.5 & 16:00:31.76 & $-$53:12:51.5 & 3.12 \\
	& G   & $-$39.7 & 16:00:31.81 & $-$53:12:52.6 & 3.13 \\
	& G   & $-$38.2 & 16:00:32.05 & $-$53:12:53.7 & 1.40 \\
	& G   & $-$37.5 & 16:00:32.01 & $-$53:12:54.7 & 4.15 \\
	G\,329.03$-$0.19$^b$ & A   & $-$46.5 & 16:00:30.29 & $-$53:12:28.6 & 16.06 \\
	& A   & $-$44.7 & 16:00:30.06 & $-$53:12:26.0 & 1.09  \\
	& B   & $-$45.5 & 16:00:30.45 & $-$53:12:13.5 & 2.99 \\
	& D   & $-$44.2 & 16:00:30.46 & $-$53:12:19.6 & 0.67 \\
	G\,329.07$-$0.30 & D   & $-$41.7 & 16:01:08.95 & $-$53:16:07.8 & 2.87 \\
	& D   & $-$40.5 & 16:01:09.33 & $-$53:16:03.5 & 0.37 \\
	& D   & $-$40.2 & 16:01:09.18 & $-$53:16:07.7 & 0.37 \\
	G\,329.18$-$0.31 & C   & $-$49.2 & 16:01:46.70 & $-$53:11:49.2 & 0.56 \\
	& E   & $-$50.2 & 16:01:47.26 & $-$53:11:43.8 & 0.70 \\
	& E   & $-$48.5 & 16:01:47.15 & $-$53:11:42.1 & 0.56 \\
	& E   & $-$47.2 & 16:01:47.33 & $-$53:11:42.3 & 2.54 \\
	& E   & $-$46.2 & 16:01:47.17 & $-$53:11:42.0 & 0.39 \\
	G\,329.47+0.50 & B   & $-$66.2 & 15:59:40.15 & $-$52:23:35.7 & 2.64 \\
	& E   & $-$69.0 & 15:59:40.66 & $-$52:23:27.5 & 0.53 \\
	& E   & $-$68.8 & 15:59:40.74 & $-$52:23:29.3 & 0.53 \\
	& E   & $-$67.3 & 15:59:40.69 & $-$52:23:28.4 & 0.36 \\
	& F   & $-$68.2 & 15:59:40.91 & $-$52:23:30.4 & 0.78 \\
	G\,331.13$-$0.24 & A   & $-$90.8 & 16:10:59.59 & $-$51:50:25.7 & 35.43 \\
	& A   & $-$86.8 & 16:10:59.74 & $-$51:50:24.5 & 4.18 \\
	& A   & $-$85.8 & 16:10:59.69 & $-$51:50:23.7 & 4.35 \\
	& A   & $-$84.8 & 16:10:59.73 & $-$51:50:24.3 & 3.28 \\
	& A   & $-$82.5 & 16:10:59.90 & $-$51:50:25.6 & 2.67 \\
	& A   & $-$80.3 & 16:11:00.00 & $-$51:50:26.7 & 0.71 \\
	& B   & $-$88.3 & 16:11:00.74 & $-$51:50:23.9 & 10.42 \\
	& C   & $-$84.0 & 16:11:00.25 & $-$51:50:25.9 & 5.08 \\
	G\,331.34$-$0.34 & A   & $-$66.5 & 16:12:26.41 & $-$51:46:18.9 & 0.60 \\
	& A   & $-$65.2 & 16:12:26.50 & $-$51:46:20.6 & 15.17 \\
	G\,331.44$-$0.18$^a$ & A   & $-$91.3 & 16:12:12.28 & $-$51:35:13.3 & 5.05  \\
	& A   & $-$89.5 & 16:12:12.23 & $-$51:35:13.0 & 1.02 \\
	& C   & $-$87.3 & 16:12:11.41 & $-$51:35:18.3 & 1.99 \\
	& D   & $-$90.5 & 16:12:12.32 & $-$51:35:10.8 & 0.28 \\
	& D   & $-$90.3 & 16:12:12.35 & $-$51:35:12.3 & 0.91 \\
	& D   & $-$89.0 & 16:12:12.33 & $-$51:35:11.0 & 0.75 \\
	& D   & $-$88.5 & 16:12:12.44 & $-$51:35:09.6 & 0.52 \\
	& D   & $-$87.5 & 16:12:12.57 & $-$51:35:08.5 & 3.21 \\
	& D   & $-$86.5 & 16:12:12.55 & $-$51:35:08.6 & 0.83 \\
	& D   & $-$86.0 & 16:12:12.39 & $-$51:35:07.4 & 0.44 \\
	& D   & $-$85.5 & 16:12:12.51 & $-$51:35:09.2 & 0.42 \\
	G\,332.30$-$0.09 & A   & $-$52.2 & 16:15:45.30 & $-$50:55:54.9 & 0.31 \\
	& A   & $-$51.7 & 16:15:45.72 & $-$50:55:55.0 & 0.64 \\
	& A   & $-$50.5 & 16:15:45.86 & $-$50:55:55.2 & 5.71 \\
	& A   & $-$49.2 & 16:15:45.78 & $-$50:55:55.3 & 19.11 \\
	& A   & $-$48.0 & 16:15:45.65 & $-$50:55:55.8 & 7.21 \\
	& A   & $-$47.2 & 16:15:45.45 & $-$50:55:55.7 & 1.23 \\
	& A   & $-$47.0 & 16:15:45.29 & $-$50:55:56.3 & 1.16 \\
	& A   & $-$46.5 & 16:15:45.45 & $-$50:55:54.3 & 0.77 \\
	& C   & $-$45.5 & 16:15:44.21 & $-$50:55:56.6 & 1.50 \\
	G\,332.60$-$0.16 & A   & $-$46.2 & 16:17:29.27 & $-$50:46:14.6 & 0.70 \\
	& A   & $-$45.5 & 16:17:29.22 & $-$50:46:12.8 & 8.52 \\
	G\,332.94$-$0.68$^a$ & A   & $-$50.0 & 16:21:19.32 & $-$50:54:13.7 & 1.52 \\
	& A   & $-$48.2 & 16:21:19.44 & $-$50:54:13.0 & 4.34 \\
	& A   & $-$46.7 & 16:21:19.32 & $-$50:54:13.8 & 1.01 \\
	G\,332.96$-$0.67$^a$ & A   & $-$50.7 & 16:21:22.93 & $-$50:52:56.7 & 0.42 \\
	& A   & $-$50.2 & 16:21:22.71 & $-$50:52:55.7 & 0.76 \\
	& A   & $-$50.0 & 16:21:22.84 & $-$50:52:57.5 & 0.49 \\
	& A   & $-$48.7 & 16:21:22.88 & $-$50:52:58.7 & 0.71 \\
	& A   & $-$48.2 & 16:21:22.84 & $-$50:52:56.9 & 2.06 \\
	& A   & $-$46.7 & 16:21:22.76 & $-$50:52:56.7 & 11.13 \\
	G\,333.03$-$0.06$^a$ & A   & $-$41.7 & 16:18:56.64 & $-$50:23:53.4 & 1.17 \\
	& A   & $-$40.3 & 16:18:56.77 & $-$50:23:54.9 & 4.27  \\
	G\,333.07$-$0.44 & - & $-$53.7 & 16:20:48.83 & $-$50:38:40.2 & 0.47 \\
	& - & $-$51.2 & 16:20:48.64 & $-$50:38:39.1 & 2.75 \\
	G\,333.13$-$0.56$^a$ & A   & $-$59.5 & 16:21:34.33 & $-$50:41:06.8 & 1.13 \\
	& B   & $-$57.2 & 16:21:35.91 & $-$50:40:49.1 & 1.45 \\
	& B   & $-$56.5 & 16:21:36.10 & $-$50:40:48.7 & 1.74 \\
	& B   & $-$56.3 & 16:21:36.33 & $-$50:40:48.3 & 1.64 \\
	& B   & $-$55.3 & 16:21:36.43 & $-$50:40:47.4 & 4.65 \\
	& B   & $-$54.0 & 16:21:36.08 & $-$50:40:48.1 & 8.69 \\
	& D   & $-$58.8 & 16:21:35.79 & $-$50:40:55.1 & 7.51 \\
	G\,333.12$-$0.43$^c$ & B   & $-$58.0 & 16:20:59.75 & $-$50:35:51.3 & 0.32 \\
	& B   & $-$56.3 & 16:20:59.71 & $-$50:35:51.7 & 0.49 \\
	& B   & $-$55.0 & 16:20:59.73 & $-$50:35:50.9 & 1.43 \\
	& B   & $-$52.3 & 16:20:59.49 & $-$50:35:50.5 & 9.31 \\
	& B   & $-$50.2 & 16:20:59.42 & $-$50:35:49.0 & 8.10 \\
	& B   & $-$49.3 & 16:20:59.63 & $-$50:35:50.2 & 1.02 \\
	& B   & $-$45.3 & 16:20:59.64 & $-$50:35:51.5 & 1.74  \\
	G\,333.13$-$0.44$^c$ & A   & $-$48.7 & 16:21:02.58 & $-$50:35:54.5 & 18.44 \\
	& A   & $-$48.0 & 16:21:02.18 & $-$50:35:50.8 & 18.85 \\
	& A   & $-$47.7 & 16:21:02.41 & $-$50:35:53.0 & 17.54 \\
	& A   & $-$47.0 & 16:21:02.63 & $-$50:35:54.0 & 19.87 \\
	& A   & $-$46.2 & 16:21:02.35 & $-$50:35:53.1 & 10.64 \\
	& A   & $-$44.5 & 16:21:02.39 & $-$50:35:52.9 & 0.49 \\
	& E   & $-$51.5 & 16:21:03.32 & $-$50:35:45.1 & 0.41 \\
	& E   & $-$50.7 & 16:21:03.16 & $-$50:35:45.9 & 1.91  \\
	& DF  & $-$49.7 & 16:21:03.73 & $-$50:35:38.8 & 12.21 \\
	& F   & $-$51.2 & 16:21:03.18 & $-$50:35:38.4 & 0.90 \\
	G\,333.16$-$0.10$^a$ & A   & $-$92.0 & 16:19:42.12 & $-$50:20:00.2 & 0.93 \\
	& B   & $-$91.0 & 16:19:42.71 & $-$50:19:55.3 & 4.74 \\
	G\,333.18$-$0.09 & A   & $-$86.8 & 16:19:45.54 & $-$50:18:32.4 & 0.80 \\
	& A   & $-$86.3 & 16:19:45.65 & $-$50:18:33.6 & 3.25 \\
	& A   & $-$85.3 & 16:19:45.96 & $-$50:18:31.9 & 0.68 \\
	G\,333.23$-$0.06 & A   & $-$89.5 & 16:19:49.52 & $-$50:15:14.1 & 0.89 \\
	& A   & $-$86.8 & 16:19:49.49 & $-$50:15:14.3 & 35.57 \\
	& C   & $-$85.0 & 16:19:51.41 & $-$50:15:14.1 & 4.90 \\
	& C   & $-$83.8 & 16:19:51.28 & $-$50:15:12.5 & 0.40 \\
	G\,333.32+0.10 & A   & $-$49.2 & 16:19:28.74 & $-$50:04:38.7 & 1.11 \\
	& A   & $-$45.5 & 16:19:29.26 & $-$50:04:43.6 & 0.38 \\
	& A   & $-$44.0 & 16:19:28.85 & $-$50:04:40.7 & 3.78 \\
	& B   & $-$48.0 & 16:19:28.14 & $-$50:04:38.7 & 0.86 \\
	& B   & $-$47.2 & 16:19:28.34 & $-$50:04:39.1 & 0.56 \\
	& B   & $-$46.7 & 16:19:27.98 & $-$50:04:42.8 & 0.56 \\
	& B   & $-$46.0 & 16:19:28.09 & $-$50:04:37.4 & 0.47 \\
	& D   & $-$44.7 & 16:19:28.62 & $-$50:04:53.5 & 0.73 \\
	G\,333.47$-$0.16 & A   & $-$45.0 & 16:21:19.94 & $-$50:09:42.0 & 3.19 \\
	& B   & $-$41.0 & 16:21:20.10 & $-$50:09:38.1 & 2.13 \\
	& C   & $-$42.3 & 16:21:20.55 & $-$50:09:55.5 & 1.80 \\
	& D   & $-$43.0 & 16:21:20.15 & $-$50:09:46.7 & 5.38 \\
	& D   & $-$40.5 & 16:21:20.42 & $-$50:09:48.8 & 0.43 \\
	& E   & $-$42.5 & 16:21:19.69 & $-$50:09:33.7 & 1.42 \\
	G\,333.56$-$0.02 & A   & $-$40.8 & 16:21:08.77 & $-$49:59:49.5 & 0.81 \\
	& A   & $-$39.5 & 16:21:08.74 & $-$49:59:49.5 & 9.35 \\
	& A   & $-$38.2 & 16:21:08.59 & $-$49:59:47.8 & 0.73  \\
	G\,335.06$-$0.42$^a$ & A   & $-$41.7 & 16:29:23.50 & $-$49:12:23.2 & 1.13 \\
	& A   & $-$41.2 & 16:29:23.46 & $-$49:12:22.5 & 0.57 \\
	& A   & $-$40.5 & 16:29:23.42 & $-$49:12:21.0 & 3.70 \\
	& B   & $-$39.7 & 16:29:22.98 & $-$49:12:32.8 & 0.31 \\
	& B   & $-$38.8 & 16:29:22.80 & $-$49:12:32.4 & 4.83 \\
	& B   & $-$37.7 & 16:29:23.04 & $-$49:12:32.1 & 3.25 \\
	& B   & $-$37.3 & 16:29:23.07 & $-$49:12:30.5 & 0.53 \\
	& B   & $-$37.0 & 16:29:23.19 & $-$49:12:29.4 & 0.47 \\
	& & & & & \\ \hline
\end{longtable} 
\begin{flushleft}
	Note: $^a$sources are not within the statistically complete sample of 6.7-GHz methanol masers\\
	~~~~~~~~$^b$ components from this target are included in the multi-pointing field image G\,329.03$-$0.20\\
	~~~~~~~~$^c$ components from this target are included in the multi-pointing field image G\,333.13$-$0.44\\
\end{flushleft}	
\twocolumn

\section{Discussion}

Throughout our discussion we will directly compare the 36- and 44-GHz methanol maser properties presented by \citet{Voronkov+14} to the current 95-GHz data. However, before doing that, it is important to highlight the differences in sample selection between the \citeauthor{Voronkov+14} study and our investigation. \citeauthor{Voronkov+14} selected almost all southern sources with previously detected class~I maser emission from the single-dish searches of \citet{Slysh+94,Valtts+00, Ellingsen05}. The target sources for these original single dish observations were highly varied, with a mix of known 6.7-GHz class II masers and H \small{II} regions. Comparatively, the targets for our investigation are based on the \citet{Ellingsen05} sample, which was more restrictive and consisted of all 6.7-GHz class II methanol masers with associated 95-GHz emission (lower detection limit of $\sim4.5$\,Jy) within the Galactic longitude range $l = 325\text{\degrees}-335\text{\degrees}$ ; $b = \pm0.53\text{\degrees}$. This means that our sample is statistically complete with a different set of selection biases present than the \citeauthor{Voronkov+14} sample. Due to all of our targets (except 333.07-0.45) being previously observed at high-resolution by \citet{Voronkov+14} we can compare all three of the observed class I transitions in our statistically complete maser target sample. Although we have 32 pointing targets total for our data, four of these pointings are associated with two class I sources (two pointings each source), therefore, in general, we will refer to a set of 30 class I sources throughout the discussion, except in cases where we are directly discussing relations between class I emission and the class II sources. 

\begin{figure}
	\includegraphics[width=\linewidth]{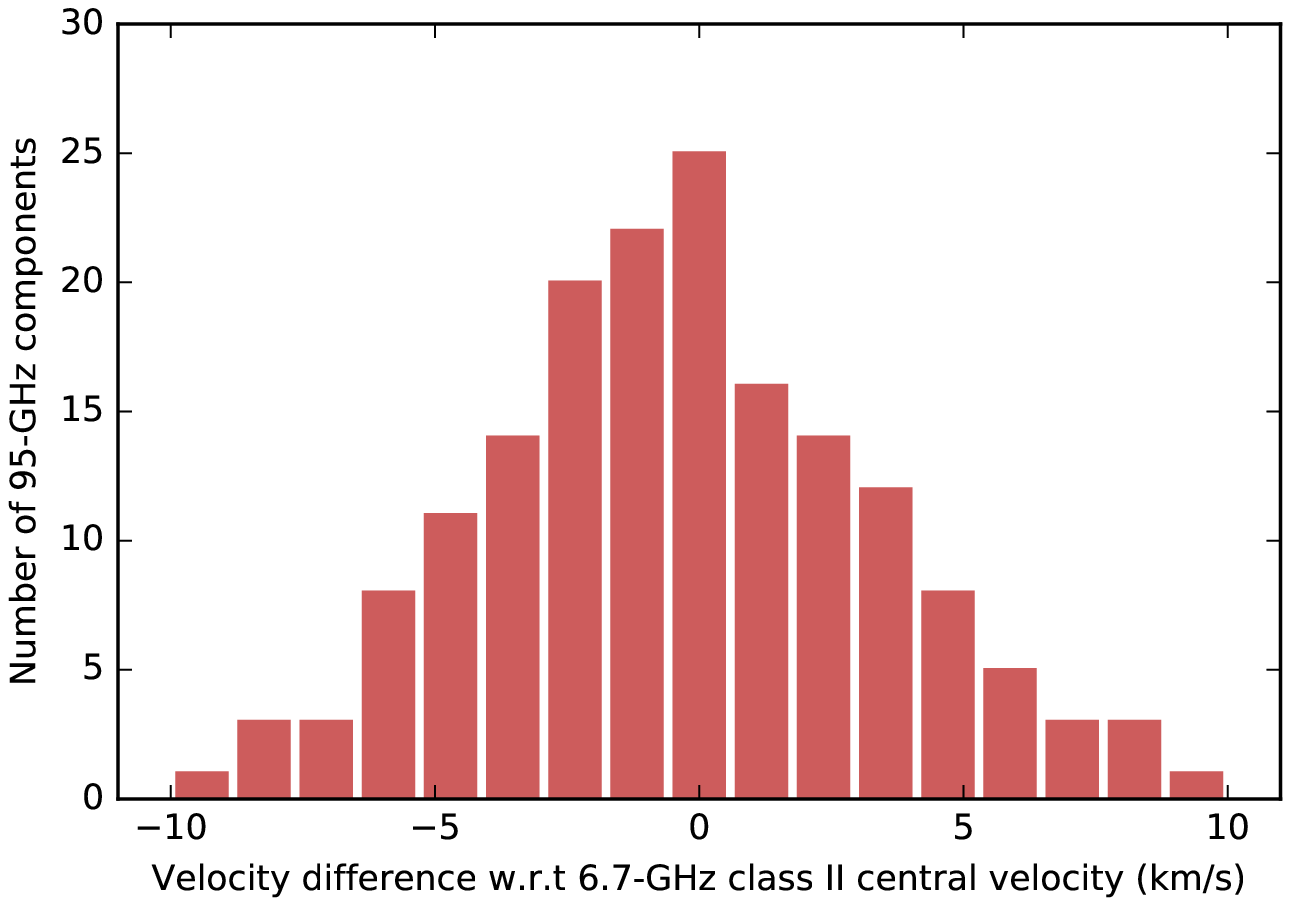}
	\caption{The histogram representing the distribution of velocity separations between the 95-GHz class I masers and the central velocity range value of the 6.7-GHz class II target sources. 
		}
	\label{fig:classI_vs_classII_velocity}
\end{figure}

\begin{figure}
	\includegraphics[width=\linewidth]{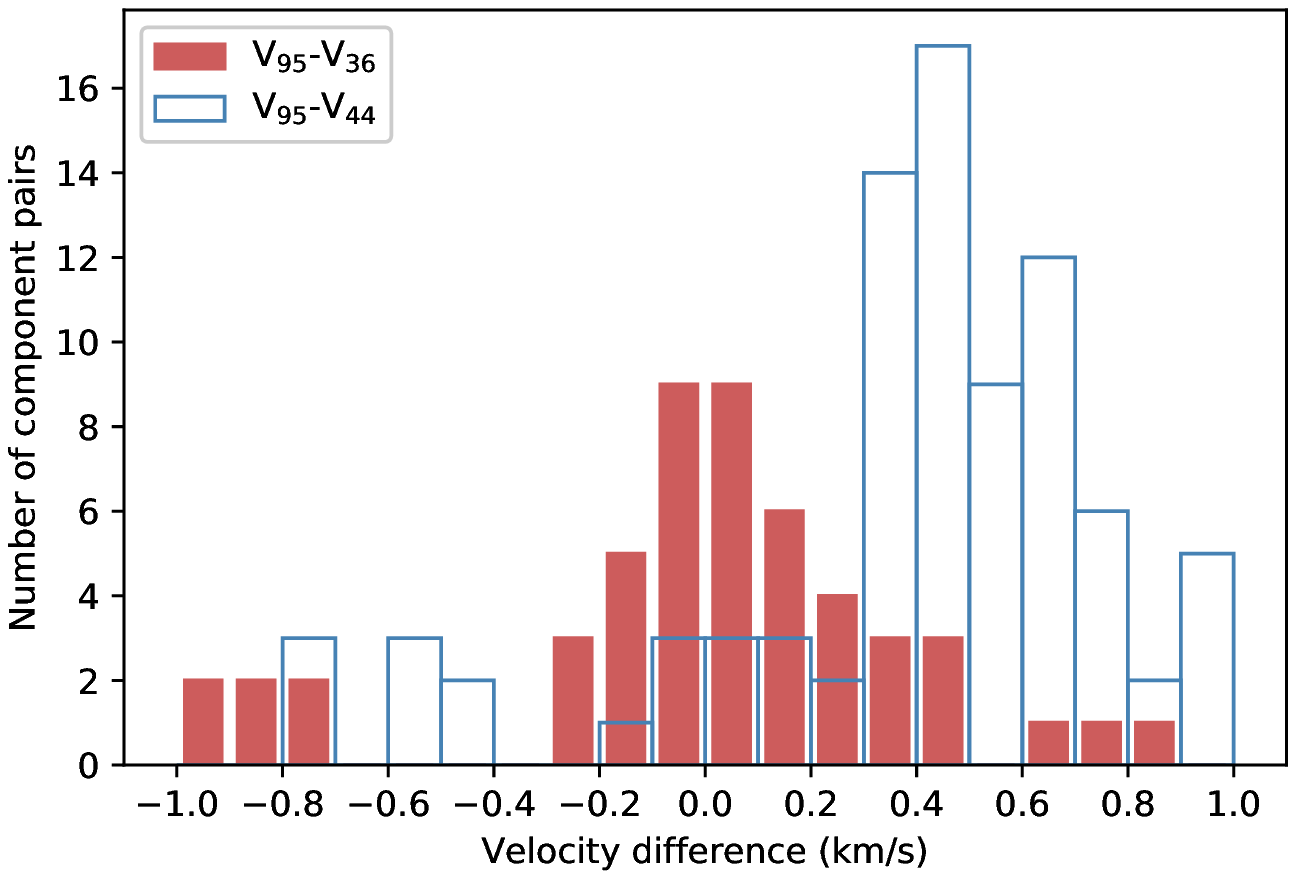}
	\caption{Histograms of velocity offset between matched 95- and 36-GHz masers (red filled) and 95- and 44-GHz masers (blue outline). }
	\label{fig:matched_vel_sep}
\end{figure}

\subsection{Diversity in class I methanol masers}

In comparison to the complex spectra observed from class II methanol masers, class I emission appears to be considerably simpler. The spectra from multiple class I transitions observed from the same masing region often appear as single components with peak emission occurring at a common velocity. This contrasts with the typical spectrum of a class II region, with multiple bright components spread across a large velocity range. However, despite the similarities in appearance of the spectra of class I transitions, our investigation combined with that of \citet{Voronkov+14} reveals significant differences between the three considered transitions. Our high-resolution observations reveal frequent differences between the physical association of the 95-GHz maser components in comparison with other phenomenon in star-formation regions, compared to the 36- and 44-GHz observations \citep{Voronkov+14}. 95-GHz masers are observed tracing ordered structures, both independently or coinciding with other class I transitions. We observe a stronger preference towards co-location with the class II maser position in the 95-GHz masers compared with the 36- or 44-GHz components. This observation is particularly interesting due to the shared optimal conditions for masing between the 95- and 44-GHz components \citep{McEwen+14, Leurini+16}. In addition, we see significant offsets in LSR velocity between matched components from the two $A^+$-type transitions (44-GHz blue-shifted compared to 95-GHz), which is not observed between the matched 95- and 36-GHz components. This diversity between transitions will be discussed further in the following sections.

\subsection{General associations}

The angular resolution of our observations allows us to determine the association of class I methanol masers and other phenomena observed in star-formation regions. The high angular resolution of the maser data combined with the \textit{Spitzer} infrared images enables us to investigate association rates between the class I emission and EGOs and other 4.5-$\mu$m excess regions \citep{Cyganowski+08}. These regions of enhanced 4.5-$\mu$m emission can result from shocked gas \citep{Debuizer+10}, therefore, we expect to see high association rates with class I methanol masers. \citet{Voronkov+14} reported association rates of $\sim$54 and $\sim$82 percent between class I emission and EGOs or any 4.5-$\mu$m excess source respectively, across their entire sample of 64 class I sources. They considered an association to exist if a class I component was within a few arcsec of the relevant region. Using this criterion for our observations, we found close spatial associations between 95-GHz class I emission and EGOs in 19 out of 30 sources ($\sim$63 percent). In 4 out of our 30 sources (G326.48+0.70, G328.24-0.55, G332.30-0.09 and G333.03-0.06), 95-GHz class I emission is associated with 4.5-$\mu$m excess sources that were not determined to be EGOs by \citet{Cyganowski+08}, two of these (G326.48+0.70 and G332.30-0.09) have associations with EGOs in addition to these other 4.5-$\mu$m excess regions. Therefore, we see 21 out of 30 sources ($\sim$70 percent) have some sort of 95-GHz maser emission associated with regions of excess 4.5-$\mu$m emission. If we instead consider any class I emission (95-GHz masers combined with 36- and 44-GHz masers reported by \citet{Voronkov+14}), we determine the following association fractions: 21 out of 30 target sources ($\sim$70 percent) contain class I masers spatially associated with EGOs as defined by \citet{Cyganowski+08} and 24 out of 30 sources (80 percent) have class I masers spatially associated with any 4.5-$\mu$m excess source. Therefore, we can conclude that in the majority of cases where both class I and class II methanol maser emission is observed, there is some component of class I maser emission is associated with a likely outflow candidate. Additionally, when comparing the association rates between EGOs and the class I transitions, we see the 95-GHz maser components have an increased association rate with confirmed EGOs compared to the other two class I transitions combined \citep{Cyganowski+08, Voronkov+14}. A review of single dish surveys of class I maser sources identified that approximately 50 percent of class I maser sources, have emission (44 or 95-GHz) within an arcmin of an outflow source (EGO or otherwise) \citep{Chen+09}, and subsequent searches for class I methanol masers targeted towards EGOs have achieved high detection rates \citep{Chen+11}. Our association rate between outflows and class I emission is significantly higher than this (80 percent when considering any class I emission in our sources) and unlike the single dish data, our high-resolution data allows us to directly confirm the presence of class I maser components within these outflow sources.

\begin{figure*}
	\begin{minipage}[b]{0.48\linewidth}
		\centering
		\includegraphics[width=\linewidth]{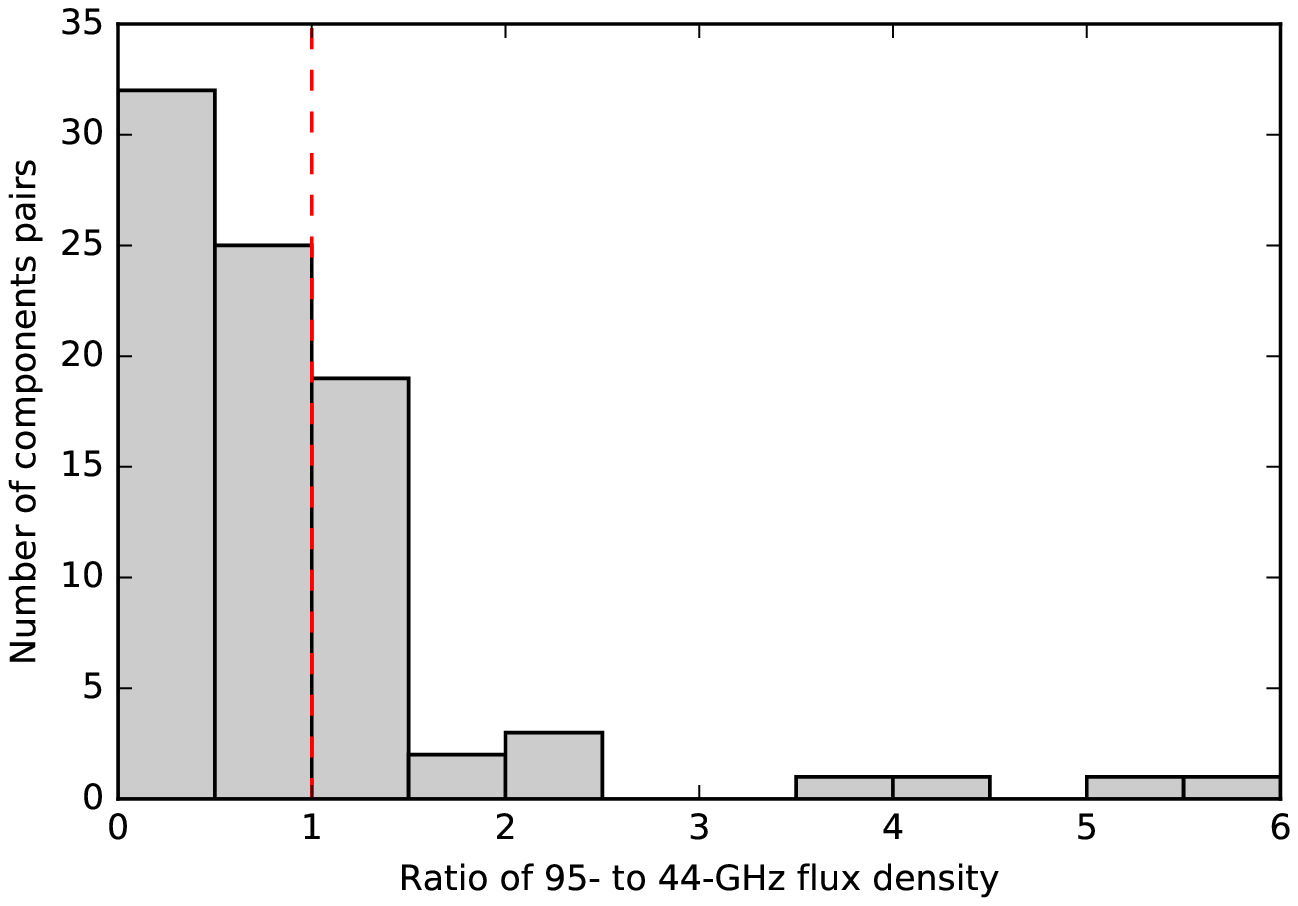}
	\end{minipage}
	\hspace{0.2cm}
	\begin{minipage}[b]{0.48\linewidth}
		\centering
		\includegraphics[width=\linewidth]{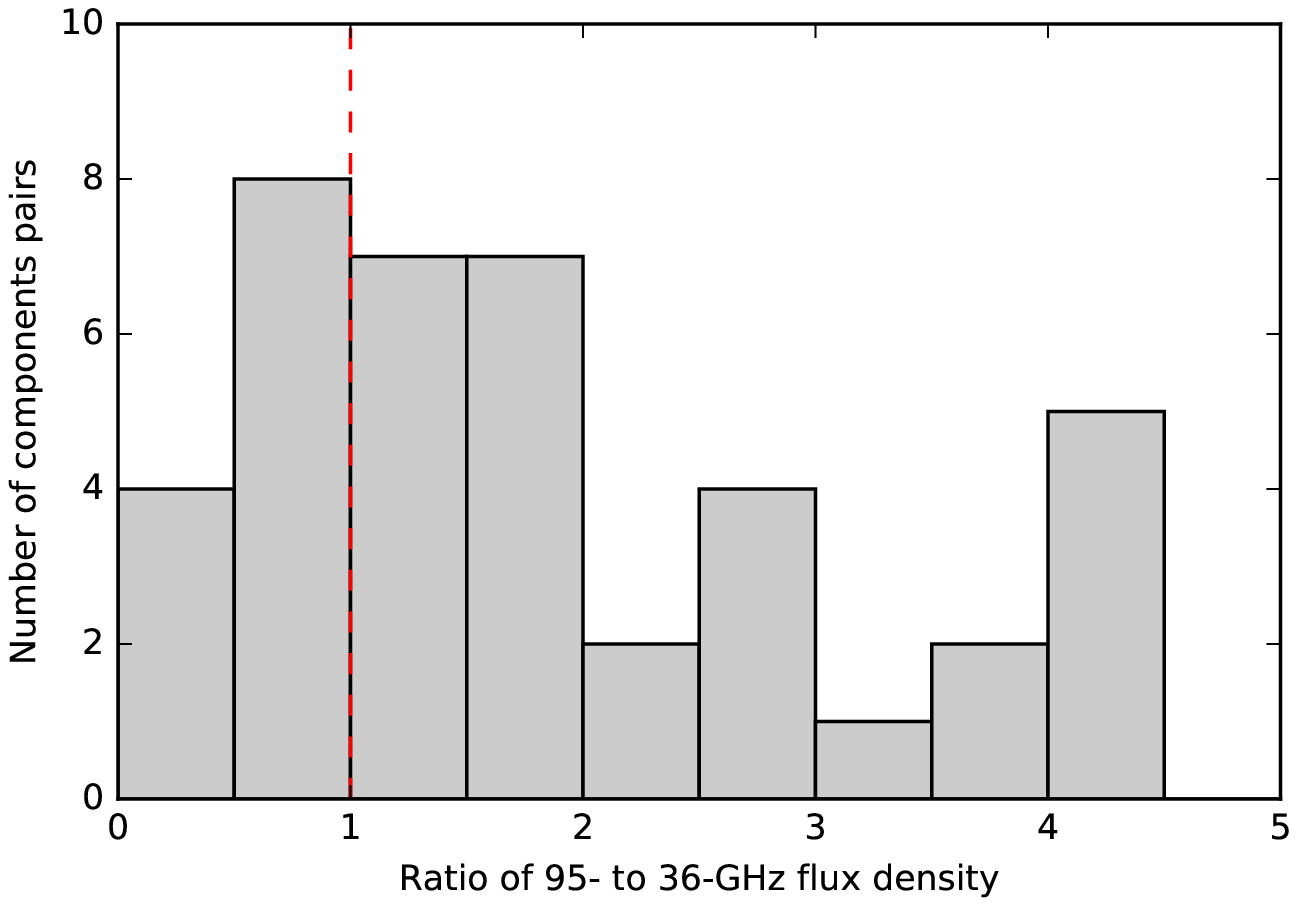}
	\end{minipage}
	\caption{Distribution of the flux density ratio of 95- to 44-GHz masers in matched pairs (left) and 95- to 36-GHz pairs (right). There are 85 and 51 matched pairs in the 95-/44-GHz and 95-/36-GHz distributions respectively. Vertical dashed line separates the pairings where 95-GHz component is brighter than the other paired class I methanol maser transition.) }
	\label{fluxratio}
\end{figure*}

There are 57 reported EGOs within the Galactic longitude range of our statistically complete sample \citep{Cyganowski+08}. 22 ($\sim$39 percent) of these EGOs have 6.7-GHz class II methanol maser associations and 14 ($\sim$25 percent) have class I methanol maser detections in one or more of the discussed class I transitions \citep{Cyganowski+08, Voronkov+14}. Indicating that EGOs with associated 6.7-GHz methanol masers make good targets for observing class I methanol masers.

Twenty of the 6.7-GHz class II methanol masers that were observed here have associated 12.2-GHz class II methanol masers \citep{Breen+12a, Breen+12b}. When considering our statistically complete sample 12 of 21 targets ($\sim$57 percent) are detected in the 12.2-GHz transition. Including all 6.7-GHz masers within the statistically complete sample our targets were drawn from (all 6.7-GHz masers, regardless of presence of 95-GHz emission) we see an association rate between the two class II transitions (6.7- and 12.2-GHz) of 60 percent. Comparing these two rates, we see that 6.7-GHz masers also displaying emission at 95-GHz do not appear to have a significantly different association rate with 12.2-GHz class II masers. Sources with emission present from 12.2-GHz class II methanol masers tend to be at a later evolutionary stage than those without \citep{Breen+10a, Breen+11}. This indicates that 95-GHz class I emission does not appear to favour these more evolved sources. 

\begin{figure}
	\captionsetup[subfigure]{justification=centering}
	\begin{center}
		\begin{minipage}[t]{1.05\linewidth}
			\begin{center}
				\subfloat[]{\includegraphics[scale=0.63]{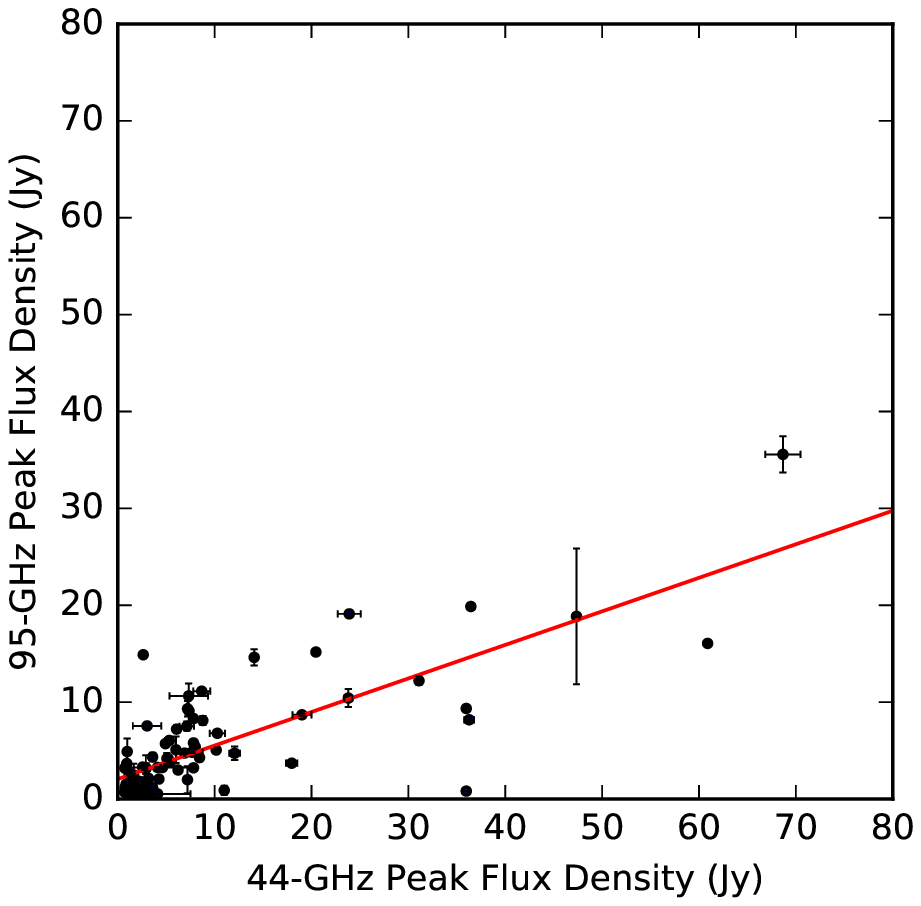}} \
				\subfloat[]{\includegraphics[scale=0.63]{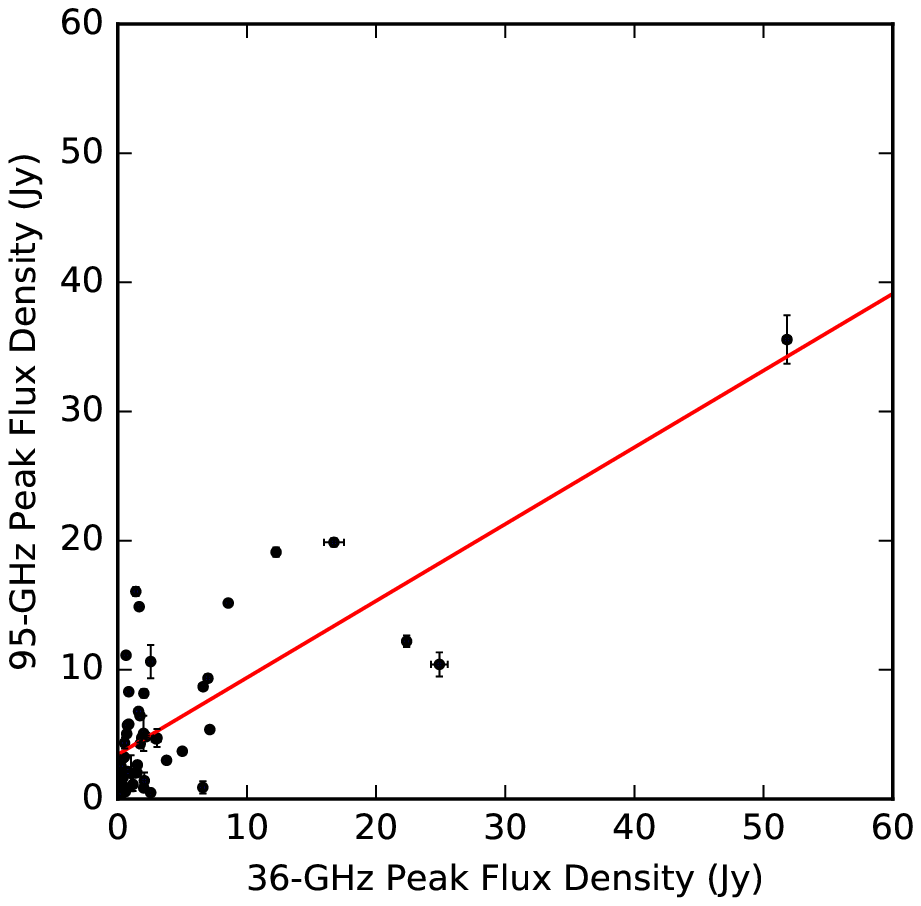}} \
				\subfloat[]{\includegraphics[scale=0.63]{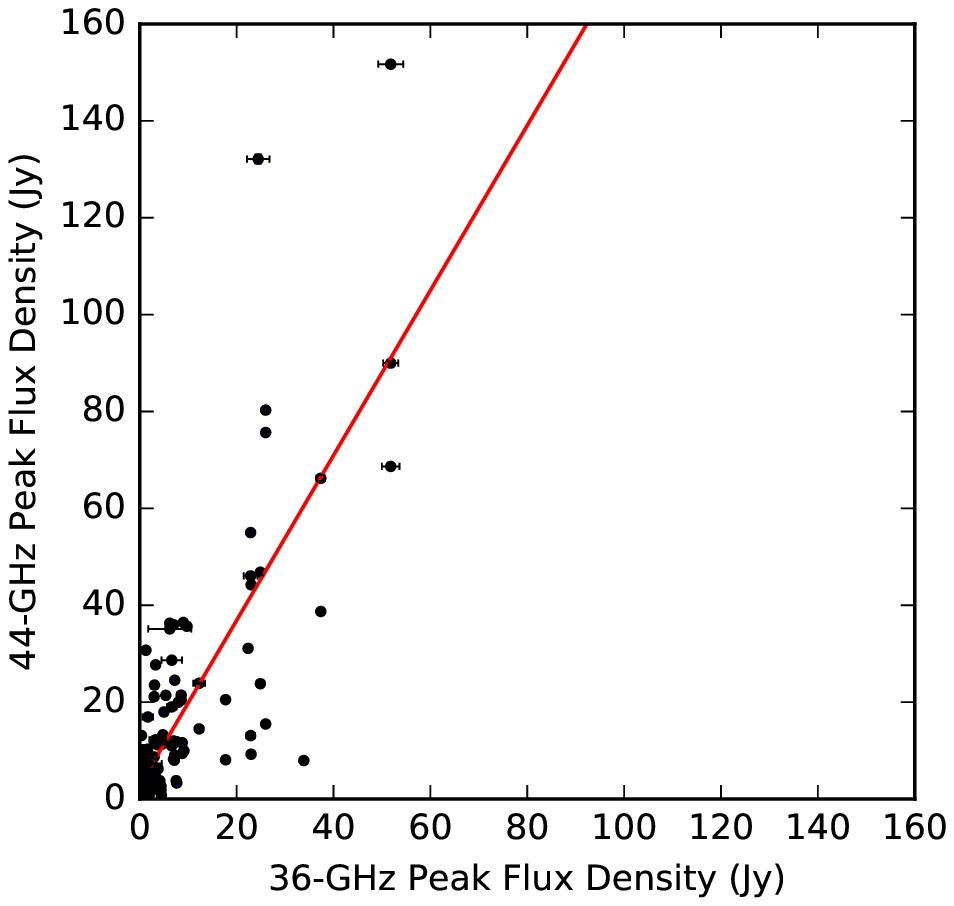}} \
			\end{center}
		\end{minipage}
	\end{center}
	\caption{Scatter plots of peak flux density between matched class I components, 95- and 44-GHz (a), 95- and 36-GHz (b) and 44- and 36-GHz (c). Line in each plot is a first-order polynomial best fit: $y=(0.35 \pm 0.10)x - (2.1 \pm 0.5)$ (a), $y=(0.59 \pm 0.17)x - (3.4 \pm 0.7)$ (b), $y=(1.70 \pm 0.49)x - (2.8 \pm 1.3)$ (c).}
	\label{flux_scatter}
\end{figure}

\subsection{Continuum association}

In addition to the observations of 95-GHz maser emission, we also searched for 3.5-mm continuum in each of our targets. We detected 3.5-mm continuum emission in 14 of our 32 pointings, with two of these pointings corresponding to the one class I methanol maser source. Therefore, 3.5-mm continuum emission is observed in 13 of the 30 ($\sim$43 percent) class I methanol maser sources (typical 3$\sigma$ upper limit is $\sim7$~mJy for non-detections). 7 of our 30 class I sources have previously reported cm-wavelength continuum emission (18-cm emission for all except G328.24$-$0.55, which has a 3-cm continuum detection) and we detect 3.5-mm continuum in all but one (G333.47$-$0.16). \citet{Voronkov+14} report an association rate of $\sim$24 percent between their class~I maser sources and cm-wavelength continuum emission. This is significantly lower than the association rate we observe between the 95-GHz and mm-wavelength continuum sources. However, it should be noted that their rate was reported as a lower limit due to the inhomogeneous nature of their continuum data with respect to their sample of sources. It is likely that mm-wavelength continuum emission, such as the observed 3.5-mm emission results from thermal dust, in contrast to cm-wavelength continuum emission which is mostly free-free emission from H \small{II} regions. This indicates that class~I sources displaying mm- and not cm-wavelength continuum emission may be at an earlier evolutionary phase.

In all of these sources we observe 95-GHz masers closely co-located (to within a few arcseconds) with the continuum emission. Additionally, in all sources where continuum emission is observed, at least some component of it was observed within several arcseconds of the class II pointing target. Considering both cm- and mm-wavelength continuum data for our sources, we obtain an association rate between class II masers and radio continuum emission ($\sim$47 percent) consistent with those previously reported \citep{Phillips+98b, Walsh+98}.

\subsection{ATLASGAL association}

The ATLASGAL survey mapped continuum emission from cold, dense dust and gas throughout the Galactic Plane. All class I methanol maser sources we have observed are within the bounds of the ATLASGAL survey. All of our class I maser regions are closely associated (within 30 arcseconds of peak emission) with ATLASGAL sources with the 95-GHz maser spots appearing to be slightly offset from the central peak of the dust emission, which is consistent with the class I masers being associated with gas outflows. We observe a positive linear correlation between the dust mass of these ATLASGAL sources, and the luminosity of the associated 6.7-GHz class II masers with a correlation coefficient ($r$) of 0.85. 

In the majority of sources that we observe strong 3.5-mm continuum emission, we observe the location of the ATLASGAL point source coinciding with the position of this continuum emission.  Additionally, we consider whether sources where 3.5-mm continuum emission or IR sources are associated with the class I maser emission, affects the dust masses of the ATLASGAL sources. This was done by using the ANOVA statistical function to determine the significance of the various categorical variables (e.g. presence of EGO, 8-$\mu$m IR source etc.) in predicting ATLASGAL dust mass. The ANOVA function is robust to the normality assumption of the continuous variable (ATLASGAL dust mass) allowing us to be reasonably confident in the output despite the skewed distribution of the dust masses. The presence of continuum emission, and 4.5-$\mu$m excess sources (whether EGO or other) do not correspond to distributions of ATLASGAL masses with statistically different means. Conversely, association with 8-$\mu$m emission does result in a statistically significant difference in the mean, with sources displaying 8-$\mu$m emission generally displaying lower dust masses.

In targets where no IR source was observed at the location of the class II maser, the ATLASGAL source likely pinpoints the location of the young star driving the methanol emission. However, in the eight sources where this is the case, we do not observe a close coincidence between the peak of the ATLASGAL emission and the location of the class II masers. We also do not observe any statistically significant difference between the luminosity of class~II masers in sources where no IR source is identified versus those with IR sources detected.

\subsection{Velocity separation}

Voronkov et al. (2014) compared relative velocities of the 36-/44-GHz class I methanol masers with the quiescent gas in the same region. Their results were consistent with the expected result that class I methanol masers have a closely related radial velocity to the systemic velocity. We follow up this analysis with a similar comparison using the detected 95-GHz masers. In the \citet{Voronkov+14} study, the velocity of the quiescent gas was approximated by the central velocity value of 6.7-GHz class II methanol maser velocity range. As the pointing targets for our observations are 6.7-GHz class II masers, we will use this same convention in our analysis. Creating a histogram (Figure \ref{fig:classI_vs_classII_velocity}) of relative velocities we see that the distribution can be reasonably approximated by a normal distribution with a mean of $-0.33\pm0.28$\,km\,s$^{-1}$ and standard deviation of $3.62\pm0.20$\,km\,s$^{-1}$. Qualitatively the histogram for the 95-GHz masers appears to have a minor blueshift asymmetry, similar to the 36- and 44-GHz transitions, however, this shift can not be considered statistically significant in our data (tested with a one sample T-Test). The ratio of negative to positive relative velocities is 1.34, this is statistically significant from unity, however, it requires a simplification of the problem to a binomial distribution and is thus not a rigorous test. Follow up observations at higher spectral resolution are required in order to conclude whether this is a real result. Unlike the 36- and 44-GHz transitions, we observed no components with extreme relative velocities, with all relative velocities falling within a range of approximately $-$10 to 10\,km\,s$^{-1}$. The 3$\sigma$ range of our results is very similar to that observed from the other class~I transitions presented by Voronkov et al. (2014). Due to the nature of our observing (covering a fixed range of -100 to 0\,\kms), some  sources such as G\,333.16--0.10, with velocities close to the edge of this range ($\sim-92$\,\kms), will not be sensitive to high velocity blue-shifted components. This same limitation is not an issue at the opposite end of our velocity range, as we do not observe any sources with systemic velocities higher than $-37$\,\kms.

\subsection{Comparison with 36-/44-GHz class I masers} \label{sec:comparison}

\citet{Voronkov+14} presented a detailed comparison between spatially associated 36- and 44-GHz methanol maser components. Combining this data with our observations allows for an analysis of the relationship between all three class I methanol maser transitions within our sample. In order to appropriately compare emission from each transition, we need to determine which maser components display emission in more than one transition. This was achieved by matching components between transitions that were co-associated both spatially and with respect to LSR velocity. 95-GHz components were matched with components from the other class I transitions if they are co-located within 1.5$''$ and separated by 1\,km\,s$^{-1}$ or less in LSR velocity. It is worth noting that while the \citeauthor{Voronkov+14} observations of the two different class I transitions occurred quasi-simultaneously, our observations were made three years prior. Therefore, any comparison between matched 95-GHz masers and 36-/44-GHz masers will have inherently more uncertainty than comparison between matched 36- and 44-GHz transitions as variability within class I methanol maser sources is not currently well understood or studied. The 36-/44-GHz masers were grouped using the same method outlined by \citeauthor{Voronkov+14}, where components from the same transition that were co-located within 3$\sigma$ (in fit uncertainty) of both position and velocity were considered a single component for the purpose of analysis. Across all 172 95-GHz components, we find matches with 36-GHz components in 51 cases ($\sim$30 percent) and matches with 44-GHz components in 85 ($\sim$49 percent). 97 out of the 172 ($\sim$56 percent) total components had a match in either of the other class I methanol maser transitions, with 39 components ($\sim$22 percent) displaying emission in all three transitions. Note that the LSR velocity of 36-GHz masers were adjusted in accordance with the rest-frequency corrections described by \citet{Voronkov+14} ($-0.215$\,km\,s$^{-1}$) before being matched with the 95-GHz masers. 

When considering 95-/44-GHz pairs we find that 57 of 85 cases ($\sim$67 percent) have higher flux density in the 44-GHz transition (left plot Figure \ref{fluxratio}). In the 95-/36-GHz matched pairs we find that the 95-GHz transition is brighter in 39 out of 51 ($\sim$76 percent) cases (right plot Figure \ref{fluxratio}). The relationship between the 95- and 44-GHz maser pairs is expected due to both of these transitions belonging to the same transition series (both are J$_0-$(J-1)$_1$A$^\text{+}$ transitions with consecutive J numbers). 

Flux density scatter plots were created between the matched components of each pair of transitions (Figure \ref{flux_scatter}). Only 36- and 44-GHz maser components from within our source sample were considered, therefore, the comparison between the paired 36- and 44-GHz masers here is a smaller sample size then that in \citet{Voronkov+14}. A first order polynomial line of best fit was determined for each of these plots. Considering the flux density scatter plot of 95- and 44-GHz matched components (plot (a) of Figure \ref{flux_scatter}), the linear fit had a slope of $0.35\pm0.10$ (and intercept value of $2.1\pm0.5$) with a modest correlation coefficient ($r$) of 0.78. This relationship is similar to the 3:1 ratio between 44- and 95-GHz flux density reported by \citet{Valtts+00}. We find reasonable coefficients of correlation ($r=0.78$ for 95-/36-GHz and 95-/44-GHz and $r=0.79$ for 44-/36-GHz) in the linear fits of all three of our transition pairs. It is important to note that these comparisons were made using the flux density peak values for each transition. This introduces velocity resolutions effects when comparing our 95-GHz data to the 36- and 44-GHz values, as the velocity resolution of our observations are approximately a factor of eight coarser. The observed relationship between frequency of association and flux density are consistent with the expected relationships between these transitions \citep{Cragg+92, McEwen+14, Leurini+16}.

When considering the velocity separation between matched class I transitions we observe a red-shift between the 95-GHz components with respect to their matched 44-GHz components in 73 out of 85 ($\sim$86 percent) pairs with a mean red-shift of 0.385\,km\,s$^{-1}$ over all pairs (see Figure \ref{fig:matched_vel_sep}). The velocity separations between the 95- to 36-GHz matched pairs has a mean of 0.005\,km\,s$^{-1}$. The distribution of the velocity separation between both pairs of transitions is approximately Gaussian. 

It is interesting to consider the number of regions where 95-GHz masers are detected which also contain emission from either, or both, of the other two class I transitions. To do this we considered the original spatial association of maser components represented by the letter symbols in our images (originally defined in \citet{Voronkov+14}). In most cases these locations have a maximum spread of approximately 4~arc~seconds, which is close enough to assume co-location in the same general environment. In a few cases 95-GHz masers were seen to be offset from an alphabetised location, in these situations they were considered their own region. This resulted in 64 groupings of 95-GHz class I masers over the 30 class I sources. In 52 of the 64 ($\sim$81 percent) clusters of 95-GHz masers, both 36- and 44-GHz masers are also present. When considering cases where only 36- or 44-GHz are co-located with the 95-GHz masers this reduces down to, $\sim$1.5 percent (1 region) and  $\sim$14 percent (9 regions) respectively, with the remaining 2 regions ($\sim$3 percent; regions in G\,328.25--0.53 and G\,329.07--0.31) only containing 95-GHz class I masers. Therefore, the vast majority of regions where 95-GHz masers are detected have 36- and 44-GHz masers in relatively-close proximity. Different environmental conditions favour different class~I methanol transitions \citep{Cragg+92, McEwen+14, Leurini+16}. For the three transitions considered here, the 44-GHz and 95-GHz emission are related and share similar optimal masing conditions, whereas the 36-GHz emission prefers denser environments. All three transitions favour temperatures $\geq50$~K, however, the optimal density range (of molecular hydrogen) for the two $A^+$ type transitions is between $10^4$ and $10^6$~cm$^{-3}$, compared with $10^5 - 10^7$~cm$^{-3}$ for the 36-GHz transition \citep{McEwen+14}. Masers from the 36- and 44-GHz transitions can be observed co-spatially due to overlapping optimal conditions for masing \citep{Voronkov+10a, Voronkov+14, Pihlstrom+11a, McEwen+14}. Therefore, as the 95-GHz masers have similar optimal conditions as the 44-GHz transition, this explains why we tend to observe spatial associations between all three transitions in regions where 95-GHz components are located (which are in general optimal for 44-/95-GHz masing, allowing 95-GHz to be observable). 

\subsection{YSO range of influence} \label{sec:yso}

Due to the nature of their pumping mechanism, 6.7-GHz class II methanol masers are located close to high-mass YSOs. This allows the positions of these class II masers to be used as an accurate indicator of the location of a YSO. As all of our target sources were 6.7-GHz class II masers, we can determine linear offsets between all our detected maser components and nearby YSOs. Kinematic distance estimates to each source were taken from \citet{Green+McClure11} or computed using similar methodology (See Table \ref{tab:observations}). Similar analysis for the 36- and 44-GHz class I emission was conducted by \citet{Voronkov+14} and concluded that 1 pc is a good estimate of the range of influence of YSOs in producing class I methanol masers. In the case of our 95-GHz observations, we have a much smaller primary beam which does not allow us to accurately determine whether the 1 pc estimate applies to the 95-GHz masers we observed. Instead, with our data we considered a range out to $0.46$ pc, this range was determined from the mean linear radius of the primary beam over all of our sources. Considering only the subset of 95-GHz components within this range limit allows for a sample with reduced bias towards detections close to the pointing centre. This reduced sample can be compared against the 36- and 44-GHz maser components, that satisfy the same criteria, in order to investigate any positional preference of 95-GHz masers towards YSOs relative to the 36- and 44-GHz masers.

\begin{figure}
	\includegraphics[width=\linewidth]{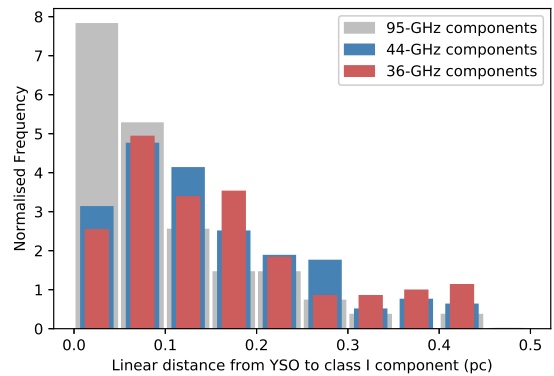}
	\caption{The normalised histogram of class I components linear separation from the 6.7-GHz (indicative of the position of the YSO). This is the reduced sample, of only components within 0.46 pc of the excitation source (see Section \ref{sec:yso}). The histogram for each frequency has been normalised so that the y-axis is the value of the probability density function for each bin, ensuring across the entire sampled range this integrates to 1. This allows comparison between the three samples of class~I masers, despite sample sizes not being equal.
	}
	\label{fig:YSO_distance}
\end{figure}

Five of our 32 target sources were excluded from this analysis. One due to a missed pointing (G\,333.56$-$0.02), and another four due to multiple close 6.7-GHz class II masers present in the field (G\,332.30$-$0.09, 333.13-0.44, G\,333.13$-$0.56 and G\,333.23$-$0.06). Sources with multiple 6.7-GHz masers are excluded as we are unable to determine which class II source the class I masers are associated with. In the remaining 27 sources we determined linear offsets for 138 components, with 127 of these falling within the range of $0.46$ pc from a YSO. Within this subset of 127 components, we observe 81 ($\sim59$ percent) with projected distances less than 0.1 pc and 107 ($\sim78$ percent) less than 0.2 pc. Therefore, we observe more than a $\sim68$ percent drop in the number of components when comparing those from between 0.1 - 0.2 pc and those within 0.1 pc of the YSO. Considering a 2-dimensional density determined as the number of components per projected area, the density of the region between 0.1 to 0.2 pc from the YSO is $\sim89$ percent lower than that of the inner region. When considering components from all three class I transitions within this same $0.46$ pc range, we must exclude three additional sources (G\,329.03$-$0.20, G\,329.03$-$0.19 and G\,333.12$-$0.43) as these have multiple 6.7-GHz class II masers within the FWHM of the 36- and 44-GHz primary beam. In this reduced sample we observe $\sim$65 percent of 95-GHz class I masers within 0.1 pc of the YSO, compared to the $\sim$37 and $\sim$39 percent observed for the 36- and 44-GHz transitions respectively (see Figure \ref{fig:YSO_distance}). Despite considering a reduced sample to alleviate bias towards 95-GHz maser components nearby to YSOs, this sample is only appropriate for comparing between the three transitions and are not indicative of the true distribution of 95-GHz maser emission in these sources. In order to accurately determine the distribution of 95-GHz maser spots with respect to the YSO, multiple pointings of each source would need to be made or an instrument with a much larger primary beam utilised.

This result, that 95-GHz masers appear to be found closer to the exciting source than the other two transitions would seem to indicate that the 95-GHz transition prefers a comparatively higher energy environment. However, the 44- and 95-GHz both share similar conditions for ideal population inversion \citep{Cragg+92, McEwen+14}, and we do not see as strong a preference for higher energy environments in the former. There is one observational factor that may be responsible for the perceived difference between these two related transitions. The 44-GHz maser transition is generally the stronger of the two transitions, so if the 95-GHz needs a higher energy environment for observable masing to occur, we may see a steeper decline in detected components when in lower energy areas, due to the sensitivity bias of our observations. This will have the effect of making 95-GHz masers seem proportionally closer to the excitation source than the 44-GHz masers. Additionally, we see no correlation between the flux density of a component (of any class I transition) and its linear separation from the associated YSO, which is consistent with the findings of \citet{Voronkov+14}. 

\subsection{Class I - Class II association at high resolution}

While the vast majority of class I methanol maser components are generally located within 1 pc of a YSO \citep{Voronkov+14}, generally, class I emission is not observed coincident with class II methanol maser features. \citet{Kurtz+04} observed examples of spatial association between 44-GHz class I and 6.7-GHz class II masers, and suggested that the mutual exclusivity between the two classes may not be as strong as previously supposed. In Section \ref{sec:yso} we observe the majority of 95-GHz emission nearby ($< 0.2$ pc) to the YSO (marked by the location of the 6.7-GHz masers) and Figure \ref{fig:classI_vs_classII_velocity} shows the distribution of velocities for our components is approximately Gaussian about the class II central velocity. Therefore, we should observe some 95-GHz maser components, both spatially associated and within the velocity range of the 6.7-GHz class II masers. Observation of components matching these two conditions does not conclusively mean masers from the two classes are associated with the same environment and could instead result simply from chance alignment.

We include in Table \ref{tab:observations} the linear offset between each class~II maser and the closest 95-GHz component coincident with the class~II maser velocity range. In 8 (of 32) class II sources we observe class I maser emission both: closer than 1 arcsecond in position and within the velocity range of the 6.7-GHz masers \citep{Caswell+11}. The median angular separation and projected linear separation of all 95-GHz components satisfying the aforementioned criteria is $0.71$ arcseconds and $0.013$ pc respectively. G\,328.25-0.53 and G\,333.03-0.06 host the two most accurately positioned components within this subset, with $3\sigma$ angular errors in their fitted positions of $<0.2$ arcseconds. The line of site distance to these sources is less than 3 kpc (see Table \ref{tab:observations}), therefore, these small angular offsets correspond to projected linear distances of less than 0.01 pc. In contrast to the 95-GHz masers, similar spatial alignment between class I maser components and 6.7-GHz masers was only observed in three sources from our sample, for both the 36- and 44-GHz transitions. Such a considerable difference between the 95- and 44-GHz masers is interesting due to their shared optimal environment conditions \citep{Cragg+92, McEwen+14}, however, it is worth noting that we are only considering 44-GHz emission from sources we have observed at 95-GHz.

From our observations it is not possible to definitively determine whether these components are anything other than aligned with the line of sight to the class~II source. In order to properly investigate coincidence between class~I and class~II masers, high resolution astrometry must be performed. However, these close angular offsets combined by the general trend that 95-GHz masers appear preferentially towards the driving source (discussed in Section \ref{sec:yso}) may indicate that masers from this transition are more strongly inverted when close to a background continuum source.

\section{Conclusions}

We present 95-GHz imaging results from 32 pointings of a statistically complete sample of 6.7-GHz class II methanol masers. We detected 95-GHz maser emission in all pointings and across all 32 pointings we detected a total of 172 95-GHz class I methanol components. Due to two cases with dual pointings, these 32 pointings represent coverage of 30 individual class I methanol maser sources. In many cases we observe similar associations between these 95-GHz maser components and ordered structures, as is seen in 36-/44-GHz components in the same sources \citep{Voronkov+14}. We observe associations between 95-GHz components and 4.5-$\mu$m excess sources in 21 out of 30 sources (70 percent) or 24 out of 30 sources (80 percent) when considering all class I methanol emission (inclusion of 36-/44-GHz components). Additionally, in 19 of 30 sources ($\sim$63 percent) we observe associations with 8.0-$\mu$m emission, increasing to 22 of 30 sources ($\sim$73 percent) with the inclusion of all class I methanol data for these sources. 3.5-mm continuum sources were detected in 13 of our 30 sources ($\sim$43 percent) and in all of these sources we see co-location (within a few arcseconds) between the continuum emission, the 95-GHz class I masers and the 6.7-GHz class II maser. 
\\ \indent When considering our statistically complete sample (all 6.7-GHz class~II sources with observed 95-GHz maser emission), we observe a 57 percent association rate with 12.2-GHz class~II masers). This is not significantly different from the 60 percent association rate observed between the two class~II transitions in the larger complete sample that our targets were drawn from. As 12.2-GHz masers generally appear at later evolutionary stages, the 95-GHz masers do not seem to prefer more evolved sources.
\\ \indent We observe components matches between the 44-GHz and 95-GHz transitions in 85 ($\sim$49 percent) of our detected 95-GHz components and matches between 36- and 95-GHz transitions in 51 ($\sim$30 percent) 95-GHz components. Only 39 of our 172 observed 95-GHz components have matching components from all three transitions. These matches were made by cross-matching every 95-GHz maser with its nearest 36- and 44-GHz masers, then filtering these matches with a limit of 1.5 arcseconds of angular separation and 1\,km\,s$^{-1}$ of LSR velocity separation. Additionally when considering broader environmental associations between the two transitions we find that in $\sim 81$ percent of regions containing 95-GHz components, 36- and 44-GHz components are also observed. Indicating that the vast majority of environments with suitable conditions for 95-GHz masing, are also suitable for 36- and 44-GHz masing. When comparing the flux densities between matched 44- and 95-GHz components we observe a similar relationship to the 3:1 ratio reported by \citet{Valtts+00}. Indicating that this relationship holds when considering high-resolution data of both transitions where individual maser components can be matched together.
\\ \indent When comparing the velocities of matched class~I components, we observe a red-shift of $0.385$ km\,s$^{-1}$ on average between 95- and 44-GHz components. No significant shift is observed between the 95- and 36-GHz matched components.
\\ \indent When considering the offset from the YSOs (position defined as location of the 6.7-GHz masers), 95-GHz maser spots appear to be preferentially closer than the other considered class~I transitions. This could be explained by the 95-GHz methanol transition being more strongly inverted when closer to a background continuum source. Additionally, we observe some 95-GHz components at very small linear offsets from class~II 6.7-GHz masers, however, high-resolution astrometry is required to determine whether any real association exists.
\\ \indent The distribution of velocity offsets between 95-GHz components and the systematic source velocity (defined by the central velocity value of the 6.7-GHz class II emission in each source) has a mean of $-0.33\pm0.28$\,km\,s$^{-1}$. We observe no high velocity features in any of our sources at 95-GHz.
\\ \indent Similar to the complementary nature of the 36- and 44-GHz components observed in our target sources \citep{Voronkov+14}, there are many cases where the 95-GHz components trace the same morphological structures in a particular source. In several of these cases, the 95-GHz components extend along paths not covered by 36- or 44-GHz components. Therefore, observations in all three transitions can be useful in determining the exact structure of particular morphological features. Our investigation rejects the notion that the class I maser transitions of the methanol species are simpler than class II, with diversity in relationships and associations being observed between closely related transitions (95-GHz and 44-GHz $A$-type).

\section*{Acknowledgements}

We thank the anonymous referee for useful suggestions which helped to improve this paper. The ATCA is part of the Australia Telescope which is funded by the Commonwealth of Australia for operation as a National Facility managed by CSIRO.  This research has made use of NASA's Astrophysics Data System Abstract Service. This research has made use of data
products from the GLIMPSE survey, which is a legacy science programme of the Spitzer Space Telescope, funded by the National
Aeronautics and Space Administration, and the NASA/IPAC Infrared Science Archive, which is operated by the Jet Propulsion
Laboratory, California Institute of Technology, under contract with the National Aeronautics and Space Administration. This research also utilised APLPY , an open-source plotting package for PYTHON hosted at http://aplpy.github.com. This research made use of Astropy, a community-developed core Python package for Astronomy \citep{astropy+13}.

\bibliography{references}

\onecolumn

\end{document}